\documentstyle[12pt]{article}
\makeindex
\newcommand{\be}{\begin{equation}}
\newcommand{\ee}{\end{equation}}
\newcommand{\bea}{\begin{eqnarray}}
\newcommand{\eea}{\end{eqnarray}}

 3
 2
 1
 1
 2
\def\rfrs#1#2{eqs.(\ref{#1})-(\ref{#2})}
\def\Rfr#1{Eq.(\ref{#1})}

\def\btab{\begin{tabular}}
\def\etab{\end{tabular}}
\def\rfr#1{eq.(\ref{#1})}
\def\dfa{\derp{\mtc{R}}{a}}
\def\dfm{\derp{\mtc{R}}{\mtc{M}}}
\def\dfog{\derp{\mtc{R}}{\og}}
\def\dfi{\derp{\mtc{R}}{i}}
\def\dfe{\derp{\mtc{R}}{e}}
\def\dfo{\derp{\mtc{R}}{\O}}
\def\cu{\rp{2}{na}}
\def\cd{\rp{1-e^2}{na^{2} e}}
\def\ctr{\rp{(1-e^2)^{1/2}}{na^{2} e}}
\def\cq{\rp{1}{na^{2}(1-e^2)^{1/2}\si}}

\def\bb{\bibitem}
\def\bar{\begin{array}}
\def\ear{\end{array}}
\def\eqi{\begin{equation}}
\def\eqf{\end{equation}}
\def\mc#1#2{\left[\matrix{#1 \cr #2\cr}\right]}

\def\mtc#1{\mathcal{#1}}

\def\ga{\gamma}
\def\d{\delta}
\def\ve{\varepsilon}

\def\th{\theta}

\def\l{\lambda}
\def\m{\mu}

\def\p{\pi}
\def\r{\rho}

\def\s{\sigma}

\def\t{\tau}

\def\f{\phi}

\def\ps{\psi}
\def\og{\omega}
\def\G{\Gamma}
\def\D{\Delta}

\def\F{\Phi}

\def\O{\Omega}

\def\ci{\cos{i}}
\def\si{\sin{i}}

\def\vass#1{\left\vert\ #1 \right\vert}
\def\rp#1#2{{#1\over#2}}

\def\derp#1#2{\rp{\partial{#1}}{\partial{#2}}}
\def\dert#1#2{\rp{d{#1}}{d{#2}}}

\def\ct#1{\cite{#1}}
\def\lb#1{\label{#1}}

\begin{document}
\begin{titlepage}
\begin{flushright}
\today\\
BARI-TH/99-353\\
\end{flushright}
\vspace{.5cm}
\begin{center}
{\LARGE Earth  tides and Lense-Thirring effect\\}
\vspace{1.0cm}
\quad\\
{Lorenzo Iorio\\
\vspace{1.0cm}
\quad\\
Dipartimento di Fisica dell' Universit\`a di Bari, via Amendola 173, 70126,
Bari, Italy\\ \vspace{1.0cm}
}
\vspace*{1.0cm}

{\bf Abstract\\}
\end{center}

{\noindent \small The general relativistic Lense-Thirring effect can be measured by inspecting a
suitable combination of the orbital residuals of the nodes of LAGEOS and LAGEOS II and the
perigee of LAGEOS II. The solid and ocean Earth tides affect the recovery of the parameter by
means of which the gravitomagnetic force is accounted for in the
combined residuals. Thus an extensive analysis of the
perturbations induced on these orbital elements by the 
solid and ocean Earth tides 
is carried out.
It involves the $l=2$ terms  for the solid tides 
and the $l=2,3,4$ terms for the ocean tides. The perigee of 
LAGEOS II turns out to be very sensitive to the $l=3$ part of the ocean
tidal spectrum, contrary
to the nodes of 
LAGEOS and LAGEOS II. The uncertainty in the solid tidal perturbations,
mainly due to 
the Love number $k_2$, ranges from $0.4\%$ to $1.5 \%$, while the ocean tides are 
uncertain at $5\%-15 \%$ level. The obtained results are used in order to
check in a preliminary way which tidal constituents
the Lense-Thirring shift is sensitive to. In particular it is tested
if the
semisecular
18.6-year zonal tide really does not affect the combined residuals.   It
turns
out that, if modeled at the level of accuracy worked out in the paper,
the $l=2,4\ m=0$ and also, to a lesser extent,
the $l=3\ m=0$ tidal perturbations cancel out .}
\end{titlepage} \newpage \pagestyle{myheadings}
\setcounter{page}{1}
\vspace{0.2cm}
\baselineskip 14pt

\setcounter{footnote}{0}
\setlength{\baselineskip}{1.5\baselineskip}
\renewcommand{\theequation}{\mbox{$\arabic{equation}$}}
\noindent

\section{Introduction}
If the Earth was perfectly spherical, according to Newton its gravitational 
potential would be  $GM_{\oplus}/r$,
 where $G$ is the Newtonian 
gravitational constant and $M_{\oplus}$ is the mass of the
Earth; the path of any artificial satellite 
 in orbit around 
it would be a Keplerian  ellipse with the Earth in one of its foci. This ellipse would 
remain fixed in inertial space and  neither its shape nor its size would change; it 
could be parameterized by means of the so called
Keplerian orbital elements [{\it Sterne}, 1960; {\it Allison and Ashby},
 1993].  
In reality, the behavior of an object orbiting the Earth is much more 
complicated because our planet is not perfectly spherical due to several 
factors. First of all, the centrifugal force caused by its diurnal rotation
 makes the Earth oblate;   its gravitational 
field, called geopotential, is no longer central but must be developed in a
 multipolar harmonic expansion \ct{kau}: 
\eqi U=\rp{GM_{\oplus}}{r}\{1+\sum_{l=2}^{\infty}\sum_{m=0}^{l}
(\rp{R_{\oplus}}{r})^{l}P_{lm}(\sin{\f})
[C_{lm}\cos{m\l}
 +S_{lm}\sin{m\l}]\}\lb{geop}\eqf where $R_{\oplus}$ is the Earth's mean 
equatorial radius, $P_{lm}$ is the associated Legendre function of degree $l$ 
and order $m$, and $\f$ and $\l$ are terrestrial latitude and east Greenwich 
longitude. $C_{lm}$ and $S_{lm}$ are the adimensional Stokes or geopotential 
coefficients [{\it Lemoine et al.,} 1998] .
Second, the shape of the Earth changes also in time due to the action of the 
solid, ocean and atmospheric tides
 which periodically lift the surface, tilt the vertical and 
redistribute the masses on our planet. This effect can 
be accounted for in a similar fashion to the static, centrifugal part of the 
geopotential given by \rfr{geop} introducing in it time
 varying-coefficients 
\ct{eanes1}.
The resulting effect of the static and dynamical
non-sphericity of the Earth is that the path of any artificial satellite
is more or less modified with 
respect to the unperturbed Keplerian ellipse. It 
is possible to account for these changes 
by  assuming that the perturbed orbit can be still
 parameterized in terms of  
Keplerian elements which
nevertheless slowly vary  in time.  This is so because the leading 
or point-mass term $U_0=GM_{\oplus}/r$ is
predominant over the perturbing terms [{\it 
Casotto}, 1989]. 
Obviously, there is also
a wide class of non gravitational effects [{\it Milani et al.,} 1987] which 
contribute to modify the
satellites' orbits; the concept
of osculating ellipse
fits also these cases and, in general, any situation in which a 
physical acceleration, small with
respect to the Newtonian monopole, acts on the test body.

It should be pointed out that the experimental accuracy of techniques 
such as the Satellite Laser Ranging (SLR) has astonishingly grown up in the 
last decade: for example, today it
 is  
possible to measure the amplitude of a periodical perturbation on the
node $\O$ of LAGEOS with an accuracy of some 
milliarcseconds (mas in the following). The accuracy for the same kind of 
measurement on the perigee $\og$ is slightly worse
(F. Vespe, private communication, 1999).  This forces the researcher who 
wants to investigate some particular physical effect by means of near-Earth's 
satellites to be very careful in accounting  for every tiny perturbation,
 of gravitational and non gravitational origin,
which would be inevitably present in the record  aliasing
 the recovered values of the 
quantities with which the
phenomenon of interest is parameterized.

This is the case for the Lense-Thirring 
drag of inertial frames [{\it Lense and Thirring}, 1918; {\it Ciufolini
and
 Wheeler}, 
1995]. It is a 
feature of the gravitational field 
generated by every rotating mass predicted by  
general relativity which consists in a tiny, secular precession  
affecting the node $\O$ and the perigee $\og$ of the  orbit of a not too far freely falling test 
body.
For LAGEOS  and LAGEOS II SLR satellites the gravitomagneic precessional rates 
in the field of the Earth amount to:
\begin{eqnarray} 
\dot \O_{LT}^{LAGEOS}&\simeq& 31 \ mas/y,\lb{letr1}\\
\dot \O_{LT}^{LAGEOS
II}&\simeq& 31.5 \ mas/y, \\
\dot \og_{LT}^{LAGEOS II}&\simeq& -57\
mas/y.\lb{letr2}
\end{eqnarray} 
The major source of uncertainty in the detection of the
gravitomagnetic shift turns out to be the mismodeling of the
first two even
zonal terms of the secular classical precessions caused by 
the oblateness of the Earth.
{\it Ciufolini} [1996]  has
proposed
a strategy which could allow to detect the Lense-Thirring effect at
$20$ $\%$ precision level by using as observable a suitable combination of
the orbital
residuals of
LAGEOS and LAGEOS II which would allow to cancel out all the static
and dynamical contributions
of the first two even zonal terms of the geopotential along with
the associated errors.      
%  If only one satellite had been employed, it would be
% impossible to detect 
%the Lense-Thirring effect on $\O$ or $\og$ because the  errors of the
%classical
%secular precessions associated to $C_{20}$ and 
%$C_{40}$ in 
%\rfr{geop} are themselves greater than the 
%predicted values of the gravitomagnetic precessions yielded by
%\rfrs{letr1}{letr2}.

Among the various sources of systematical errors which may affect the recovery of the 
Lense-Thirring effect the Earth tides [{\it Cartwright}, 2000] play  
a relevant role.

The  calculations presented here,
% of the tidal perturbations on the orbital elements of the LAGEOSs
in the context of  the gravitomagnetic experiment,
can be viewed as a preliminary sensitivity test of the Lense-Thirring
shift to the tidal perturbations.
Moreover, in view of the wide application of the two LAGEOS in many fields
of space science they could turn out to be useful for other experiments.

An evaluation from
first principles of the amplitudes, the periods and the initial phases
of the tidal perturbations on the nodes of LAGEOS and LAGEOS II and
the perigee of LAGEOS II is
useful because:\\
$\bullet$ It allows to point out which tidal constituents
the Lense-Thirring shift is mainly sensitive to,
so that people can focus the researches on them\\
$\bullet$ It allows to get insight in how to update the orbit
determination
softwares. In this way the impact on the time series of those tidal constituents which will turn out to be more
effective could be reduced along with the number
of the parameters to be included in the least-squares solution\\
$\bullet$ For a given observational time span  $T_{obs}$ it allows to
find those constituents whose periods  are
longer than it and consequently may act like superimposed
bias (e.g. the 18.6-year and 9.3-year zonal tides and the $K_1\ l=3\
p=1\ q=-1$
ocean diurnal tide on the perigee of LAGEOS II) corrupting the detection
of the gravitomagnetic shift.
In these cases, if we know their periods and initial
phases we
could try to fit and remove them from the time series\footnote{If the
period of the disturbing harmonic, assumed to be known, is shorter than
the
time series length
the perturbation can be reliably viewed as an empirically fit
quantity. But if its
period
is longer
than the time series length it is not possible to fit and remove the
harmonic
without removing also the true linear trend of interest.} without
affecting also the trend of interest or, at least, it
should be possible to assess the mismodeling level induced by them on the
detection of the Lense-Thirring trend so to include also these estimates in the final error
budget\\
$\bullet$ It can be used as starting point for numerical simulations of
the time series in order  to check, e. g.,
the impact of the diurnal ($m$=1) and semidiurnal ($m$=2) tidal
perturbations
(not
canceled out by the Ciufolini' s observable), as done in a further paper 
[{\it Pavlis and Iorio}, 2001].
%Moreover, apart from the semisecular
%18.6-year tide which should be canceled out, there are other
%long period tidal perturbations affecting the proposed
%combined
%residuals, like the $K_1\ l=3\ p=1\ q=-1$, which,
%according to the length of the time series, could manifest as aliasing
%trends.

Regarding the claimed cancellation of the perturbations induced by
the  first two even zonal
static and dynamical terms of the geopotential by the proposed combined
residuals, it has been  
checked,  as
far as the tides are concerned, if it really occurs and in this case at
which
level of
accuracy  it takes place, and if this feature
can be extended to some other tidal constituents by updating and extending the
calculations performed with the nominal amplitudes in [{\it Ciufolini et 
al.}, 1997 pag. 2714 eq.(20), eq.(21)]      

The paper is organized as follows:                           
In Sec. 2
the perturbative amplitudes on $\O$ and $\og$ of
LAGEOS and LAGEOS II due to the solid tides are worked out. The calculation 
account for 
the frequency-dependence of Love numbers $k_2$ and the anelasticity of the Earth's 
mantle. For some selected tidal lines\ -the most effective in perturbing
the 
satellites' orbits-\ also the effect of Earth's 
rotation and  ellipticity, accounted for by the coefficients
$k_2^{+}$, are investigated. 
%The mismodeled perturbative amplitudes of
%the solid tidal spectrum  are consequently compared
%to the
%gravitomagnetic
%perturbations over 4 years in order to check
%which solid tidal constituents, over such temporal span, are mismodeled
%at 
%$1\%$ level of the
%Lense-Thirring effect. 
In Sec. 3  analogous calculations for the ocean tidal spectrum are performed.
%The mismodeled perturbative amplitudes of the ocean tides  are then
%compared to the
%gravitomagnetic perturbations
%over 4 years in order to check which ocean tidal
%constituents, over such temporal span, are mismodeled at
%$1\%$ level of the
%Lense-Thirring effect.
In Sec. 4 
the mismodeled perturbative amplitudes of
the solid and ocean tidal spectrum  are  compared
to the
gravitomagnetic
shift over 4 years in order to check
which tidal constituents, over such temporal span, are mismodeled at
$1\%$ level of the
Lense-Thirring effect.
Subsequently, the obtained results are employed to check extensively the
feasibility of the 
cancellation of the zonal tidal constituents by means of the 
formula proposed by Ciufolini. 
Sec. 5 is devoted to the conclusions.

The data employed in the present analysis are in the following table in which
$P$ is the orbital period and $P[X]$ is the period of the 
Keplerian element $X$:
\begin{eqnarray} 
a_{LAGEOS}&=&12,270, km\\
a_{LAGEOS II}&=&12,163, km\\
e_{LAGEOS}&=&0.0045\\
e_{LAGEOS II}&=&0.014\\
i_{LAGEOS}&=&110, deg\\
i_{LAGEOS II}&=&52.65, deg\\
P_{LAGEOS}&=&0.1566, days\\
P_{LAGEOS II}&=&0.1545, days\\
P[\O]_{LAGEOS}&=&1,043.67, days\\
P[\O]_{LAGEOS II}&=&-569.21, days\\
P[\og]_{LAGEOS}&=&-1,707.62, days\\
P[\og]_{LAGEOS II}&=&821.79, days\\
\end{eqnarray}  
\section{Effect of the solid Earth  tides on the nodal and apsidal lines of 
LAGEOS and LAGEOS II}   
The free space potential [{\it Dahlen,} 1972; {\it Smith,} 1974; {\it Wahr}, 1981b; {\it Melchior,} 1983; {\it Wang,} 1997] by means
of which the Earth\footnote{It is assumed that the Earth is endowed with a solid inner core, a
fluid outer
core, and a solid mantle capped by a thin continental crust, without oceans and
atmosphere.  Substantially, the Earth
is thought as a set of coupled harmonic oscillators showing a variety of
resonant frequencies -the normal modes- which can be excited by the external
forcing constituents of the tide generating potential [{\it Wahr,} 1981a].}
responds to the tide generating potential $\F({\bf r})$ [{\it Doodson}, 1921;
{\it Cartwright and Tayler}, 1971;
{\it Cartwright and Edden}, 1973;
{\it Buellsfeld}, 1985; {\it Tamura}, 1987; {\it Xi}, 1987
; {\it Hartmann and Wenzel}, 1995; {\it Roosbeek}, 1996], at $r=a$ for a near circular orbit satellite, is
given by:\eqi \f=
\rp{GM_{\oplus}}{R_{\oplus}}\sum_{l=2}^{\infty}\sum_{m=0}^{l}
\rp{1}{a}({\rp{R_{\oplus}}{a}})^{l}\sum_{p=0}^{l}
\sum_{q=-\infty}^{+\infty}F_{lmp}(i)G_{lpq}(e)S_{lmpq}(\omega,\ \O,
{\mtc{M}},\ \th).\lb{freep}\eqf In \rfr{freep}   $F_{lmp}(i)$m
and $G_{lpq}(e)$ are the inclination and eccentricity functions respectively
[{\it Kaula,} 1966]
and the $S_{lmpq}$ is given by:
\eqi S_{lmpq}(\omega,\ \O,
{\mtc{M}},\ \th)=A_{lm}\sum_{f}H_{l}^{m}
k^{(0)}_{lm}(f)\cos{(\s t+\ps_{lmpq}-\d_{lmf})}.\lb{somg}\eqf
In \rfr{somg} $\og,\ \O$ and $\mtc{M}$ are the argument of perigee, the
longitude of ascending node and the mean anomaly, respectively, of the
satellite, $\th$ is the Greenwich sidereal time, while $A_{lm}$ is given by:
\eqi A_{lm}=\sqrt{\rp{2l+1}{4\p}\rp{(l-m)!}{(l+m)!}}.\lb{cosho}\eqf
In \rfr{cosho} the Condon-Shortley phase factor $(-1)^{m}$ has been neglected in order to compare the results of this Section to those by [{\it
Bertotti and Carpino}, 1989]. The phase of the generic harmonic in \rfr{somg} is constituted by
three terms. The first one,
$\s t$,
is for every constituent of the tide generating potential
a linear combination of the Mean Lunar Time $\t$ and the mean ecliptical
longitudes of the astronomical motions of the Moon and the Sun [{\it Dronkers
,} 1964] so that:
\eqi \s\equiv 2\p f= j_1\dot \t+j_2\dot s+ j_3 \dot h+j_4 \dot p+ j_5\dot N^{'}+ j_6
\dot p_s.\lb{tre}\eqf  Since $\t=\th-s$ \rfr{tre} becomes:
\eqi \s= j_1\dot \th+(j_2 -j_1)\dot s+ j_3 \dot h+j_4 \dot p+ j_5\dot N^{'}+ j_6
\dot p_s.\lb{tre'}\eqf The coefficients $j_k,\ k=1,...,6$ are small integers
which can assume negative, positive or null values. They are
 arranged in the so called Doodson number:
\eqi j_1(j_2+5)(j_3+5).(j_4+5)(j_5+5)(j_6+5),\eqf by means of which every tidal
constituent is named. In it the integer $j_1$
classifies the tides in long period or zonal ($j_1=0$),
 diurnal or tesseral ($j_1=1$) and
semidiurnal or sectorial ($j_1=2$). According to the Doodson notation, the   
sum $\sum_{f}$ over the tidal constituents in \rfr{somg} stands for the sum
over the integers $j_1,...j_6$.

The second quantity appearing in the phase of \rfr{somg} is:
\eqi\ps_{lmpq}=
(l-2p)\omega+(l-2p+q){\mtc{M}}+m(\O-\th)\lb{quattro}.\eqf
The parameter:
\eqi
k^{(0)}_{lm}(f)=\sqrt{{k_{lm}^{R}(f)}^2+{k_{lm}^{I}(f)}^2},\lb{loven}\eqf
is the modulus of the Love number
$k$ [{\it Love}, 1926;  {\it Mathews et al.,} 1995] for a static, spherical Earth. In general,
it is defined as the ratio of the free space
gravitational potential $\f({\bf r})$, evaluated at the Earth' s equator, and the tide generating
potential $\F({\bf r})$ calculated at the mean equatorial radius. 
 If the rotation and nonsphericity of the Earth are accounted for, there
are different values of $k$ for any $l,m$ and $f$. In general, they are evaluated by convolving, in the
frequency domain, the tide
generating potential
with the transfer function for a non rigid Earth. There are several theoretical calculations for the Love numbers
$k$ [{\it Wahr},
1981b; {\it Wang,} 1994;
{\it Mathews et al.,} 1995; {\it McCarthy}, 1996; {\it Dehant et al.,} 1999]. They differ in the
choice of the Earth' s interior
models [{\it Gilbert and Dziewonski,} 1975; {\it Dziewonski and Anderson,} 1981]
adopted for
the calculation of the transfer function and the departure from   
elasticity of the mantle` s behavior.
Particularly important is also if they account for, or not, the normal modes in the diurnal band.
  In \rfr{loven} the IERS  conventions [{\it McCarthy,} 1996]
have been used for $k^{(0)}_{lm}(f)$:
 \eqi k_{lm}^{I}(f)=Im\ k_{lm} + \d k_f^{anel},\eqf
\eqi k_{lm}^{R}(f)=Re\ k_{lm} + \d k_f^{el},\eqf
where  $Im\ k_{lm}$ and
 $Re\ k_{lm}$ are the frequency-independent parts of Love numbers,
and
$\d k_f^{anel}$,
$\d k_f^{el}$ are the frequency-dependent corrections. They are important in the diurnal band, through
$\d k_f^{el}$, and in the zonal band with $\d k_f^{anel}$ due to the anelasticity of the mantle.

The factor $\d_{lmf}$ is the phase lag of the response of the solid Earth with
respect to the tidal constituent of degree $l$, order $m$ and circular frequency
$\s$ induced by the anelasticity in the mantle [{\it Varga,} 1998]: \eqi
\tan{\d_{lmf}}=\rp{k^{I}_{lm}(f)}{k^{R}_{lm}(f)}.\eqf Note that if
$k_{lm}^{I}(f)=0$,
also $\tan{\d_{lmf}}=0$.
It may be important to know these parameters. Indeed, the initial phases
of some semisecular tidal constituents, like the 18.6-year
tide, consist entirely of them and the astronomical
longitudes of the Sun and the Moon. Indeed \rfr{quattro} vanishes for
the even ($l-2p=0$) zonal ($m=0$) long period
($l-2p+q=0$) perturbations. For
a time span $T_{obs}$ shorter than their periods they could be fitted and
removed from the time series provided that their
periods and initial phases are exactly known. The same remarks will hold
for the ocean tides.

The quantities $H_l^m(f)$ appearing in \rfr{somg}
are the coefficients of the harmonic expansion of the
tide generating potential. They contain the lunisolar ephemerides
information and define the modulus of the vertical shift $V/g$ in the
equipotential surface at the Earth's surface for $r=R_{\oplus}$
with respect to the case
in which only proper Earth's gravity is considered; $g$ is the acceleration of
gravity at the equator. The values used in the
present
calculation are those recently released by {\it Roosbeek}  [1996]. Their
accuracy is of the order of $10^{-7}$ m. In order  to use them in place of the
coefficients of {\it Cartwright and
Edden} [1973], which have a different
 normalization, a multiplication for a suitable conversion factor
[{\it McCarthy}, 1996] has
been performed.

The \rfr{freep}
 is the dynamical, non central part of the geopotential due to the
response of the solid Earth to the forcing lunisolar tidal
solicitation.  In the linear Lagrange equations of perturbation theory for the
rates of change of the
Keplerian elements of a satellite it  plays
 the role of the perturbative function $\mtc{R}$. The latter is the non
central
part of the total mechanical energy of the satellite, with the sign reversed,
according to the geophysical convention.
The Lagrange equations  are [{\it Kaula}, 1966]:
\begin{eqnarray}\dert{a}{t}&=& \cu \ \dfm ,\lb{sei}\\
\dert{e}{t}&=& \cd \ \dfm - \ctr \ \dfog ,\\ 
\dert{i}{t}&=& \ci \cq \ \dfog-\nonumber \\
           & & - \cq \ \dfo ,\\
\dert{\O}{t}&=& \cq \ \dfi ,\\
\dert{\og}{t}&=&- \ci \cq \ \dfi +\nonumber \\
 & &+ \ctr \ \dfe ,\\
\dert{\mtc{M}}{t}&=&n- \cd \ \dfe - \cu \ \dfa .\lb{sette}\end{eqnarray}
In it $n=(GM_{\oplus})^{1/2}a^{-3/2}=2\p/P$ is the mean motion of the satellite.
It is an observed fact that the secular motions are the dominant perturbation
in the elements $\og,\ \O,\ {\mtc{M}}$ of
geodetically useful satellites. So, taking as constants $a,\ e,\
i$ and consider as linearly variable in time
$\og, \O,\ {\mtc{M}}$ and $\th$,
besides $\s t$, the following expressions, at first order, may be
worked out:
%%%%%%%%%%%%%%%%%%%%%%%%%%%%%%%%%%%%%%%%%%%%%%%%%%%
$$\D\O_f=\rp{g}{na^2\sqrt{1-e^2}\sin{i}}
\sum_{l=0}^{\infty}\sum_{m=0}^{l}
({\rp{R}{r}})^{l+1}\times$$\eqi\times A_{lm}
\sum_{p=0}^{l}
\sum_{q=-\infty}^{+\infty}\dert{F_{lmp}}{i}G_{lpq}\rp{1}{f_p}
k^{(0)}_{lm}H_{l}^{m}\sin{\ga_{flmpq}},\lb{nodo}\eqf
%%%%%%%%%%%%%%%%%%%%%%%%%%%%%%%%%%%%%%%%%%%%%%%%%%5 
$$\D\og_f=\rp{g}{na^2\sqrt{1-e^2}}
\sum_{l=0}^{\infty}\sum_{m=0}^{l}
({\rp{R}{r}})^{l+1}\times$$\eqi\times A_{lm}\sum_{p=0}^{l}
\sum_{q=-\infty}^{+\infty}[
\rp{1-e^2}{e}F_{lmp}\dert{G_{lpq}}{e}-
\rp{\cos{i}}
{\sin{i}}\dert{F_{lmp}}{i}G_{lpq}]\rp{1}{f_p}
k^{(0)}_{lm}H_{l}^{m}\sin{\ga_{flmpq}},\lb{perig}\eqf
%%%%%%%%%%%%%%%%%%%%%%%%%%%%%%%%%%%%%%%%%%%%%%%%%%%%%%%%%%%
where:
\eqi\ga_{flmpq}=(l-2p)\omega+(l-2p+q){\mtc{M}}+m(\O-\th)+\s t-\d_{lmf},
\eqf
\eqi f_p=(l-2p)\dot\omega+(l-2p+q)\dot{\mtc{M}}
+m(\dot\O-\dot\th)+\s.\lb{pertfr}\eqf
Only the perturbations whose periods are much
longer than those of the orbital satellites motions,
 which, typically,
amount to a few hours, are relevant for many geophysical and astronomical
 applications.
This implies that in the summations of \rfrs{nodo}{perig}
only those
 terms
in which the rate of the mean anomaly does not appear must be retained,
i. e. the condition: \eqi
l-2p+q=0\lb{undici}\eqf
must be fulfilled. Moreover, if  the effect of Earth's diurnal rotation,
which could introduce periodicities of the order of 24 hours, is to be
neglected, one  must retain only those terms in which  the non negative
 multiplicative coefficient $j_1$ of the Greenwich sidereal time in $\s t$
coincides to  the order $m$ of the tidal constituent considered:
 in this way in $f_p$ the contributions of
$\dot \th$ are equal and opposite, and cancel out. 
 With these bounds on $l,\ m,\ p$ and $q$ the circular
frequencies of the perturbations of interest  become:
\eqi  f_p=\dot \G_f+(l-2p)\dot \og+m\dot \O\lb{dodici}\eqf in which:
\eqi
\dot \G_f=(j_2-m) \dot s+j_3 \dot h+j_4 \dot p+j_5 \dot N^{'}+j_6 \dot p_s.
\eqf
In the performed calculation
only the degree $l=2$ constituents have been considered due
to the smallness of the $k_{3m}^{(0)}$ and $H_{3}^{m}(f)$.
For $l=2$, $p$ runs from $0$ to $2$, and so, in virtue of the condition
$l-2p+q=0$,
$q$ assumes the values $-2,\ 0,\ 2$. From an inspection of the table of the
eccentricity functions $G_{lpq}(e)$
in [{\it Kaula}, 1966] it turns out that $G_{20-2}=G_{222}=0$, while
$G_{210}=(1-e^2)^{-3/2}$. For this combination of $l,\ p$ and $q$ the relation
$l-2p=0$ is fulfilled:  the
frequencies of the perturbations are, in this case, given by:\eqi f_p =\dot
\G_f +m\dot \O\lb{tredici}.\eqf

In Tab.\ref{uno}, Tab.\ref{due} and Tab.\ref{trec} 
 the  results for the nodal lines of LAGEOS 
and LAGEOS II, and the apsidal line of LAGEOS II are shown;
  since the 
observable quantity for $\og$ is
 $ea\dot \og$ \ct{ciuff},   the calculation for the perigee of  
LAGEOS, due to the notable smallness of the  eccentricity of its orbit, have
 not been performed.
The tidal lines for which the analysis was performed have
 been chosen  also in order to make a comparison with the perturbative 
 amplitudes of the ocean tides 
based on the results of EGM96 gravity model
which will be shown in the next Section.

The results presented in Tab.\ref{uno}, Tab.\ref{due} and
 Tab.\ref{trec} can be compared to those of {\it Dow} [1988] and Carpino
 [{\it Bertotti and Carpino}, 1989]. In doing so it must be kept in mind that both these authors have 
neglected not only the contribution of $k_{lm}^{+}$ but also the anelasticity and the frequency dependence of the 
Love numbers $k_{lm}^{(0)}$. Moreover, Dow includes in his analysis, in a not entirely clear  manner, also the 
indirect influences of the oblateness of the Earth [{\it Balmino,} 1974; {\it Dow,} 1988; {\it Casotto}, 1989]. 
Obviously, in their analyses the LAGEOS II is not present since it was launched 
only in 1992. Another important factor 
to be considered is the actual sensitivity in measurements of $\O$
and $\og$, in the sense that the eventual discrepancies between the present results and 
the other ones must be not smaller than the experimental error in the Keplerian 
elements if one wishes to check the theoretical assumptions behind the different 
models adopted.  
 Carpino has analyzed
 the inclination and the node of LAGEOS only.  His value  
for the zonal 18.6-year tide is $-1087.24$ mas, while  Tab.\ref{uno} gives $-1079.38$ mas; 
the difference amounts to $7.86$ mas, the $0.72$ $\%$ of the ``elastic'', 
frequency-independent Carpino's value. Considering that for the node $\O$ of 
LAGEOS the present accuracy is of the order of the mas, $7.86$ mas 
could be in principle detected, allowing for a discrimination between the 
different models adopted in the calculation. For the $K_1$ tide, one of the 
most powerful constraints in perturbing the satellites' orbits,  Tab.\ref{uno}
quotes  $1744.38$ mas
against $2144.46$ mas of Carpino's result; the gap is $400.08$ mas, the $18.6$ $\%$ of 
Carpino. In the sectorial band, the present analysis quotes  for the $K_2$ $-92.37$ mas while  Carpino has
$-97.54$ mas; there is a difference of  $5.17$ mas, the $5.3$ $\%$ of the
Carpino's value. It must pointed out that there are other tidal lines for which the difference 
falls below the mas level, as is the case for the 9.3-year tide. As it could be 
expected, the major differences between the present ``anelastic'',
 frequency-dependent 
calculations and
 the other ones based on a single, real value for the Love number 
$k_2$ lie in the diurnal band: in it the contribution of anelasticity is not 
particularly relevant, but, as already pointed out, the elastic part of 
$k_{21}^{(0)}$ is strongly dependent on  frequencies of the tidal spectrum.
May be interesting to note that when the calculation  have been  repeated
with the same 
value $k_2=0.317$ adopted by Carpino,  his results have been obtained again.

Up to now the effects induced by the Earth's flattening and the Earth's 
rotation have been neglected. If they have to be analyzed,  it is necessary to 
take  in the free space potential' s expansion the part [{\it Wahr,} 1981b]:
\eqi Re\ g\sum_{l=2}^{\infty}\sum_{m=0}^{l}\sum_f H_{l}^{m}(f)(\rp{R}{r})^{l+3}
k_{lm}^{+}(f) \ Y_{l+2}^{m}(\f,\ \l)\lb{kappa+}\eqf and work out it in the same 
manner as done for the spherical nonrotating Earth's contribution. In applying 
the standard transformation to the orbital elements of [{\it Kaula}, 1966] one has to 
substitute everywhere $l$ with $l+2$. So the equations for the perturbations on 
the node and the perigee become:
$$\D\O^{+}_f=\rp{g}{na^2\sqrt{1-e^2}\sin{i}}
\sum_{l=0}^{\infty}\sum_{m=0}^{l}
({\rp{R}{r}})^{l+3}A_{l+2 \ m}H_{l}^{m}\times $$
\eqi \times k^{+}_{lm}\rp{1}{f_p}\sum_{p=0}^{l}
\sum_{q=-\infty}^{+\infty}\dert{F_{l+2\ mp}}{i}G_{l+2\ pq}
\sin{\ga_{fl+2\ mpq}},\lb{juun+}\eqf
%%%%%%%%%%%%%%%%%%%%%%%%%%%%%%%%%%%%%%%%%%%%%%%%%%5
$$ \D\og^{+}_f=\rp{g}{na^2\sqrt{1-e^2}}
\sum_{l=0}^{\infty}\sum_{m=0}^{l}
({\rp{R}{r}})^{l+3}A_{l+2\ m}
k^{+}_{lm}H_{l}^{m}\rp{1}{f_p}\times $$
$$\times\sum_{p=0}^{l}
\sum_{q=-\infty}^{+\infty}[
\rp{1-e^2}{e}F_{l+2\ mp}\dert{G_{l+2\ pq}}{e}- $$
\eqi -\rp{\cos{i}}
{\sin{i}}\dert{F_{l+2\ mp}}{i}G_{l+2\ pq}]\sin{\ga_{fl+2\ mpq}}.\lb{du+}\eqf
The  corrections  induced by the Earth's flattening and rotation, due to the 
smallness of $k^{+}_{lm}$, have been calculated only for those tidal lines 
which have resulted to be the most powerful in perturbing the node and the 
perigee of LAGEOS and LAGEOS II, i.e. the
zonal 18.6-year tide and the $K_1$.
Adopting the values quoted in [{\it Dehant et al.}, 1999] for $k^{+}_{lm}$ and
 the \rfr{juun+} and \rfr{du+} it is possible to obtain the results summarized 
in Tab.\ref{k+}. The quoted values, with the exception of the node of
 LAGEOS II, fall below the mas level resulting, at the 
present, undetectable.

Concerning the mismodeling of the solid tidal perturbations of the nodes of LAGEOS and LAGEOS II and the
perigee of LAGEOS II, particularly important for the gravitomagnetic experiment, by
inspecting the analytical expressions of their perturbations given in \rfrs{nodo}{perig} it is possible to note that almost all
quantities entering in them are very well determined with the possible exception of the inclination $i$ and the Love number $k$.
Concerning the inclination $i$, by assuming  $\d i=0.5$ mas [{\it Ciufolini}, 1989], we have calculated $\vass{\derp{A(\O)}{i}}\d i$ and
$\vass{\derp{A(\og)}{i}}\d i$
for the 18.6-year tide,  $K_1,\ $and $S_2$ which are the most powerful tidal constituents in perturbing LAGEOS and LAGEOS II orbits, as can be
inferred from Tab.\ref{uno}-Tab.\ref{trec}. The results are of the order
of $10^{-6}$ mas, so that we can neglect the effect of
uncertainties in the inclination determination. About the Love numbers $k_2$, we have assessed the uncertainties in them by
calculating
for 
certain tidal constituents the factor $\d k_2/{k_2}$; ${k_2}$ is the average on the values released by the most reliable
models and $\d k_2$ is its standard deviation. According to the recommendations of the Working Group of Theoretical Tidal Model of the
Earth Tide Commission (http://www.astro.oma.be/D1/EARTH$\_$TIDES/wgtide.html), in the diurnal band we have 
chosen the values released by 
{\it Mathews et al.} [1995], {\it McCarthy} [1996] and the two sets by {\it Dehant et al.} [1999]. For 
the zonal and sectorial bands we have included also 
the results of {\it Wang} [1994]. The uncertainties calculated in the Love numbers $k_2$ span from 0.4$\%$ to 1.5$\%$ for the tides of
interest.
%These results have been obtained in order to calculate
%the mismodeled amplitudes of the solid tidal perturbations
%$\d\O^{I},\ \d\O^{II}$ and $\d\og^{II}$; they have been
%subsequently  compared with the gravitomagnetic precessions over 4
%years $\D\O^{I}_{LT}=124$ mas, $\D\O^{II}_{LT}=126$ mas
%and $\D\og^{II}_{LT}=-228$ mas. In Tab.\ref{mismo1} we have quoted
%those tidal lines whose mismodeled perturbative
%amplitudes are greater than $1\%$ of the gravitomagnetic
%perturbations. It
%turns out
%that only the 18.6-year tide and $K_1$ exceed this cutoff.
\section{Ocean tidal perturbations on the nodal and apsidal lines of
LAGEOS and LAGEOS II}
The effects of the ocean tides [{\it Defant,} 1961; {\it Neumann et al.,} 1966;
{\it Pekeris and Akkad}, 1969; {\it Hendershott and Munk},
1970; {\it Hendershott},
1972; 1973; {{\it Schwiderski,} 1980; 1981; {\it Gill,} 1982, {\it Zahel,} 1997], for a
 constituent of given frequency $f$,
 can globally accounted for by means of the
potential:
\eqi  U_f= 
\sum_{l=0}^{\infty}\sum_{m=0}^{l}\sum_{+}^{-}({\rp{R_{\oplus}}{r}})
^{l+1}A_{lmf}^{\pm}
\sum_{p=0}^{l}
\sum_{q=-\infty}^{+\infty}F_{lmp}(i)G_{lpq}(e)
{\mc{\cos{\ga_{flmpq}^{\pm}}}{\sin{\ga_{flmpq}^{\pm}}}}
^{l-m\ even}_{l-m\ odd},\lb{ciol}\eqf
in which:
\eqi \ga_{flmpq}^{\pm}=(l-2p)\omega+(l-2p+q){\mtc{M}}+m(\O-\th)\pm(\s
t-\ve_{lmf}^{\pm})\lb{gamma},\eqf
\eqi A_{lmf}^{\pm} =
 4\p G R_{\oplus} \r_{w}\rp{(1+k^{'}_{lf})}{2l+1}C_{lmf}^{\pm}.\lb{quattr}\eqf
In \rfr{ciol}
expressions like
 $\sum_{+}^{-}A^{\pm}\cos{(a\pm b)}$ mean
$A^{+}\cos{(a+b)
}+A^{-}\cos{(a-b)}$. The sign $+$ refers to the progressive (westwards) waves
while the sign $-$ is for the waves moving from West to East. Moreover,
$\r_{w}$ denotes the water density [{\it McCarthy,} 1996], $k^{'}_{l}$ are the load
Love numbers [{\it Farrell,} 1972] accounting for the ocean loading [{\it Jentzsch,} 1997], the  $C_{lmf}^{\pm}$ are the ocean tidal
coefficients
of EGM96 gravity
model [{\it Lemoine  et. al.}, 1998] and
$\ve_{lmf}^{\pm}$ are the phase shifts due to hydrodynamics of
the oceans.      

The equations for the tidal ocean perturbations may be worked out as already
done for the solid Earth tides in the previous Section.
At first order, one obtains:
%%%%%%%%%%%%%%%%%%%%%%%%%%%%%%%%%%%%%%%%%%%%%%%%%%%%%%%%%
$$ \D\O_f=\rp{1}{na^2\sqrt{1-e^2}\sin{i}}
\sum_{l=0}^{\infty}\sum_{m=0}^{l}\sum_{+}^{-}
({\rp{R}{r}})^{l+1}A_{lmf}^{\pm}\times $$
\eqi \times\sum_{p=0}^{l}
\sum_{q=-\infty}^{+\infty}\dert{F_{lmp}}{i}G_{lpq}\rp{1}{f_p}
\mc{\sin{\ga_{flmpq}^{\pm}}}{-\cos{\ga_{flmpq}^{\pm}}}^{l-m
\ even}_{l-m\  odd},\lb{nodoc}\eqf   
%%%%%%%%%%%%%%%%%%%%%%%%%%%%%%%%%%%%%%%%%%%%%%%%%%5
$$ \D\og_f=\rp{1}{na^2\sqrt{1-e^2}}
\sum_{l=0}^{\infty}\sum_{m=0}^{l}\sum_{+}^{-}
({\rp{R}{r}})^{l+1}A_{lmf}^{\pm}\times $$
\eqi\times\sum_{p=0}^{l}\sum_{q=-\infty}^{+\infty}[
\rp{1-e^2}{e}F_{lmp}\dert{G_{lpq}}{e}-\rp{\cos{i}}
{\sin{i}}\dert{F_{lmp}}{i}G_{lpq}]\rp{1}{f_p}
\mc{\sin{\ga_{flmpq}^{\pm}}}{-\cos{\ga_{flmpq}^{\pm}}}^{l-m
\ even}_{l-m\ odd},\lb{pgoc}\eqf
%%%%%%%%%%%%%%%%%%%%%%%%%%%%%%%%%%%%%
in which:
\eqi f_p=(l-2p)\dot \og+(l-2p+q)\dot {\mtc{M}}+ m(\dot \O-\dot \th)\pm
 {\s}.\lb{fr}\eqf
It should be noted that the frequencies of the perturbations given by
 the \rfr{fr}
are different, in general, from the frequencies of the solid Earth tidal
perturbations given by \rfr{pertfr}.
 While for the solid tides the diurnal modulation due to
$\dot \th$ cancels out automatically if one considers those terms
 in which $j_1=m$,
for the ocean tides, in general, this does not happen because
of the presence of the Eastwards waves due to the non equilibrium pattern of
the ocean tidal bulge.
Long periodicities can be obtained considering those combinations of $l,\ p$
and $q$ for
which $l-2p+q=0$ and retaining  the Westward prograde terms with
$j_1=m$. Only
in this way in \rfr{fr} the contributions of
$\dot \th$ are equal and opposite, and cancel out.
 With these bounds on $l,\ m,\ p$ and $q$ the frequencies of the perturbations
of interest  become:
\eqi  f_p=\dot \G_f+(l-2p)\dot \og+m\dot \O\lb{ff}.\eqf
It is worthwhile noting that the frequencies $f_p$
are identical to those of solid tidal perturbations, so that the
satellites cannot distinguish one effect from the other.
For $l=2$, as for the solid tides, $l-2p=0$ holds.
The
frequencies of the perturbations are, in this case, given by:\eqi f_p=
\dot \G_f+m
\dot\O\lb{ft}.\eqf
If $l$ is odd the situation is different because now all the terms with $p$
running from
0 to $l$ must be considered and  $q$ is
different from zero: $q=2p-l$. So, the rates of the perturbations include the
contribute of $\dot \og$.

The \rfrs{nodoc}{pgoc} have been adopted in order to compute the amplitudes
of the ocean tidal  perturbations on the Keplerian elements $\O,\ \og$, ${\mtc{M}}$ and $i$
for LAGEOS and
LAGEOS II. The calculation have been performed, considering  
only the progressive waves,  for the following tidal lines:\\
$M_m\ (065.455),\ S_a\ (056.554),\ M_f\ (075.555),\ S_{sa}\ (057.555);\\
K_1\ (165.555),\ O_1\ (145.555),\ P_1\ (163.555),\ Q_1\ (135.655);$\\
$K_2\ (275.555)$, $M_2\ (255.555)$, $S_2\ (273.555)$, $N_2\ (245.655)$,
$T_2\ (272.556)$.\\
For each of these tidal constituents the following terms have been calculated:
\\
$ l=2,\ p=1,\ q=0$ because the eccentricity
functions $G_{lpq}(e)$ for $p=0,\ q=-2$ and $p=2,\ q=2$ vanish.\\
$l=3,\ p=1,\ q=-1$ and $l=3,\ p=2,\ q=1$ because $G_{30-3}$ and
 $G_{333}$ are not quoted in (Kaula 1966) due to their smallness: indeed, the
$G_{lpq}(e)$ are proportional to $e^{|q|}$.\\
$ l=4,\ p=2,\ q=0$ because the other admissible combinations of $l,\ p$
and $q$ give rise to negligible eccentricity functions. So, also for $l=4$ the
condition
$l-2p=0$,
in practice, holds and the constituents of degree $l=2$ and $l=4$ generate
detectable perturbations with identical periods.
Similar analyses can be found in [{\it Christodoulidis}, 1978; {\it Dow}, 
1988]. When Christodoulidis 
performed his study, which is relative to only five constituents, neither
the LAGEOS 
nor the LAGEOS II were in orbit, while Dow   has 
sampled the tidal spectrum for LAGEOS in a poorer manner with respect to this 
study
in the sense that, for each constituent,
only the terms of degree $l=2$ have been 
considered with the exception of the  $K_1$ whose $l=4$ contribution has been 
also analyzed.  Moreover, when these work have been realized there were a few 
coefficients $C_{lmf}^+$ available with relevant associated errors.

It should be pointed out that the values obtained for the 
coefficients $C_{lmf}^+$ are, in general, 
biased by  the effects of the anelasticity of the solid Earth's mantle and by 
all other phenomena which have not been explicitly modeled in the nominal
background models used [{\it Lemoine et al.,} 1998]. 
For example, in the values recovered for the $S_a$ are included climatological 
effects which have not gravitational origin; in the $S_2$ coefficients are also 
included the variations of the atmospheric pressure due to the atmospheric 
tides. 

In Tab.\ref{otto}, Tab.\ref{nove} and Tab.\ref{dieci}
the present results for the nodes $\O$ of LAGEOS and 
LAGEOS II, and the 
argument of perigee $\og$ for LAGEOS II are quoted.
% and LARES satellites.
        
From an accurate inspection of Tab.\ref{otto}, Tab.\ref{nove}
% and Tab.\ref{nolar}
 it is possible to note that, for the nodes $\O$, only the even 
degree terms give an appreciable contribute; the $l=3$ part of the spectrum is  totally 
negligible. This so because $\D \O_f$ is proportional to the 
 $G_{lpq}(e)$ functions which, in turn, are, in general,  proportional to 
$e^{|q|}$:
in this case  the eccentricity 
functions used are $G(e)_{31-1}=e(1-e^2)^{-5/2}=G(e)_{321}$.

Among the long period zonal tides, the Solar annual tide $S_a$ (056.554) exerts 
the most relevant action on the nodes, with an associated percent 
error in the amplitudes of $6.7$ $\%$. It is interesting to compare for this 
tidal line the ocean tidal 
perturbations of degree $l=2$ $A^{ocean}_{LAGEOS}(\O)=-20.55$ mas,
 $A^{ocean}_{LAGEOSII}(\O)=37.7$ mas, with those due to the solid Earth tides 
$A^{solid}_{LAGEOS}(\O)=9.96$ mas, $A^{solid}_{LAGEOSII}(\O)=-18.28$ mas. The 
ocean amplitudes amount to $204$  $\%$ of the solid ones, while for the other 
zonal constituents they vary from $9.9$ $\%$ for the $M_m$ to the $18.5$ $\%$ of the 
$S_{sa}$. This seems to point toward that the recovered value  of 
$C_{20f}^{+}$ for the $S_a$ is biased by other climatological effects than the 
tides; indeed, for all the other tidal lines, zonal or not, 
the ocean tidal perturbations of degree $l=2$ amount to almost $10$ $\%$ of the corresponding solid 
Earth tides perturbations.
 
The terms of degree 
$l=2$ of the tesseral tides $K_1$ (165.555) and $P_1$ (163.555) 
induce very large 
perturbations on the node of LAGEOS 
% and LARES
 and, to a lesser extent,  of 
LAGEOS II; Tab.\ref{otto} and Tab.\ref{nove} quote $156.55$ mas, $-11.49$ mas for the former and $-35.69$ mas, 
$-3$ mas for the latter. The associated percent
errors are $3.8$ $\%$ and $8.1$ $\%$, 
respectively, for the two constituents.
For the terms of degree $l=4$, which have the same periods of those of degree 
$l=2$, the situation is analogous: Tab.\ref{otto} and Tab.\ref{nove} quote $4.63$ mas and $-0.32$ mas for LAGEOS 
and $41.58$ and $0.7$ mas for LAGEOS II. The percent errors associated 
with $K_1$ and $P_1$, for $l=4$, are $3.9$ $\%$ and $8$ $\%$ respectively. 
For all the tesseral lines investigated the ocean tidal perturbations of degree 
$l=2$ are, in 
general, the $10$ $\%$ of the solid tidal perturbations.

Among the sectorial tides, the most relevant in perturbing the nodes of LAGEOS  
satellites, on large temporal scales, is the $S_2$ (273.555): Tab.\ref{otto} and Tab.\ref{nove} quote $9.45$ mas ($l=2$) and $15.08$ 
mas ($l=4$) for LAGEOS,  and $6.87$ mas ($l=2$) and $6.79$ mas ($l=4$)
 for LAGEOS II. The associated percent errors are $3.9$ $\%$ for 
$l=2$ and $5.2$ $\%$ for $l=4$. 
The $S_2$ ocean perturbation of degree $l=2$ amounts to  nearly $5$ $\%$ of the corresponding solid tide. 

About  the  zonal 18.6-year and
 9.3-year tides,  recently both Starlette and LAGEOS SLR satellites passed their 19th 
year in orbit and this span of time is now adequate to get reliable information 
about these tides [{\it Cheng}, 1995; {\it Eanes}, 1995]  due to their slow frequencies they can 
be adequately modeled in terms of the equilibrium theory through the $H_l^m$ 
coefficients and a complex Love number accounting for the anelasticity of the 
mantle. So, concerning them the results quoted for the solid tides can be considered 
adequately representative.   

In Tab.\ref{dieci}
% and Tab.\ref{pelar}
 the  amplitudes of the perturbations
 on the 
argument of perigee $\og$ for LAGEOS II 
% and LARES
 are quoted. 
For this orbital element the factor: \eqi 
\rp{1-e^2}{e}F_{lmp}
\dert{G_{lpq}}{e}-\rp{\cos{i}}{\sin{i}}\dert{F_{lmp}}{i}G_{lpq}\eqf
to which $\D \og_f$ is proportional makes the contributions of the $l=3$ terms 
not negligible.
For the even degree terms the situation is quite similar to that of $\O$ in the 
sense that the most influent tidal lines are the $S_a,\ K_1,\ P_1$ and $S_2$.\\ 
Once again, among the long period tides the $S_a$ 
exhibits a characteristic behavior. Indeed, its $l=3$ contributions are much 
stronger than those of the other zonal tides. This fact could be connected to 
the large values obtained in its $C_{30f}^{+}$ coefficient and people believe 
that it partially represents north to south hemisphere mass transport effects 
with an annual cycle nontidal in origin.
The $l=3$ terms present, in general, for the perigee of LAGEOS II a very interesting spectrum also for the tesseral and 
sectorial bands:  for LAGEOS II
% and LARES 
there are lots of tidal lines which 
induce, on large temporal scales, very relevant perturbations on $\og$, with periods of the order of 
an year or more. Above all, it  must be quoted the effect of $K_1$ 
line for $p=1,\ 
q=-1$:  the perturbation induced amounts to -1136 mas  
with period of -1851.9 days. These values are comparable to the 
effects induced by the solid Earth tidal constituent of degree $l=2$ on the node 
$\O$ of LAGEOS. By comparing the  $l=2$ terms 
with the corresponding solid tides, it can be noted that also for the perigee the 
proportions are the same already seen for the nodes.

About the mismodeling of  the ocean tidal orbital perturbations, which are the worst
determined part of the spectrum of the Earth tidal response,
the major source of uncertainties in their amplitudes resides in the EGM96 coefficients $C^{+}_{lmf}$
and the ocean loading
parameters $k^{'}_{l}$. Concerning the ocean loading,  {\it Pagiatakis} [1990] in a first step has recalculated $k^{'}_{l}$ for an elastic,
isotropic and
non-rotating Earth: for $l<800$ he claims that his estimates differ from those by {\it Farrell} [1972], calculated with the same hypotheses, of
less
than $1\%$. Subsequently, he added to the
equations, one at a time, the effects of anisotropy, rotation and dissipation; for low values of $l$ their effects on the results of
the calculations amount to
less than
1$\%$. By inspecting Tab.\ref{otto}-Tab.\ref{dieci} it has been decided to calculate $\vass{\derp{A(\og)}{k^{'}_{l}}}\d k^{'}_{l}$ of the
perigee of LAGEOS II for $K_1\ l=3\ p=1$. First, we have calculated mean and standard deviation of the values for $k^{'}_{3}$ by Farrell
and Pagiatakis obtaining $\d k^{'}_{3}/k^{'}_{3}=0.9\%$, in according to the estimates by Pagiatakis. Then, by assuming in a pessimistic way
that the global effect of the departures
from these symmetric models yield to  a total $\d k^{'}_{3}/k^{'}_{3}=2\%$,
we have obtained $\d\og^{II}=5.5$ mas. Subsequently, for this constituent and for all other tidal lines we have calculated the effect of the
mismodeling of $C^{+}_{lmf}$ as quoted in EGM96.
%In Tab.\ref{mismo2} we compare the so obtained mismodeled ocean tidal
%perturbations to those generated over 4 years
%by the Lense-Thirring effect. It turns out that the perigee of LAGEOS II
%is more sensitive to the mismodeling of the ocean part of the Earth
%response to the tide generating potential.
%In particular, the whole effect of $K_1\ l=3\ p=1\ q=-1$ is relevant with
%a total $\d\og=
%\vass{\derp{A(\og)}{C^{+}_{lmf}}}\d C^{+}_{lmf}+
%\vass{\derp{A(\og)}{k^{'}_{l}}}\d k^{'}_{l}$ of 64.5
%mas
%amounting to 28.3 $\%$ of $\D\og_{LT}$ over 4 years.
%------------------------------------------------------ 
\section{The influence of the tides on the detection of the Lense-Thirring
drag}
The theoretical general relativistic expressions for the
 gravitomagnetic precessions of the node and
the perigee of a test particle in the field of a massive rotating central
body -the Lense-Thirring effect- are
given by:
\bea
\dot\O_{LT}&=&\rp{G}{c^2}\rp{2J}{a^{3}(1-e^2)^{3/2}},\label{lt1}\\      
\dot\og_{LT}&=&-\rp{G}{c^2}\rp{2J}{a^{3}(1-e^2)^{3/2}} 3\ci,\label{lt2}\eea
in which:\\
$J$\ \ \ angular momentum
 of the central rotating body, $g\ cm^{2}s^{-1}$. For
the Earth its value is $5.9\cdot 10^{40}\ g\ cm^{2}s^{-1}$.\\
$c$\ \ \ speed of light in vacuo, $cm\ s^{-1}$.

According to [{\it Ciufolini}, 1996; {\it Ciufolini et al.} 1997; 1998], it should be possible to detect the
Lense-Thirring shift  at a 20 $\%$ level of accuracy through the
formula: \eqi
\d\dot\O^{I}+c_1\d\dot\O^{II}+c_2\d\dot\og^{II}=60.05\times
\m_{LT},\lb{resciu}\eqf where
$c_1\simeq 0.295,\ c_2\simeq -0.35$, $\m_{LT}$ is an adimensional  scale
parameter  which is 1 in general relativity and 0 in
Newtonian mechanics and $\d\O^{I},\ \d\O^{II},\ \d\og^{II}$ are the
residuals, in mas,
of the nodes
of LAGEOS and LAGEOS II and the perigee of
LAGEOS II calculated, e.g., with the aid of the NASA-Goddard software
GEODYN II.
%   The residuals account for
%all the mismodeled or unknown physical
%phenomena acting upon the satellites' orbital elements. 
In the analyses performed with the real data [{\it Ciufolini et al.,} 1997; 1998] the
Lense-Thirring parameter $\m$ has
been purposely left out the dynamical models of GEODYN II by setting
$\m=0$ so that the residuals can entirely account for it.
%Among the various mismodeled phenomena
%affecting \rfr{resciu}  there are also  the
%Earth solid and ocean tides,
%and our task is to use
%the results obtained in the previous Sections in order to assess, in the
%more refined manner as
%possible,
%the influence of the even zonal tides on the recovery of $\m_{LT}$.
General relativity predicts for \rfr{resciu} a secular trend with a
slope of 60.05 mas/y if it is calculated by means of \rfrs{lt1}{lt2}.
\Rfr{resciu} has
been
recently investigated in [{\it Ciufolini et al.}, 1998] for a 4
years time span.

In view of a refinement of the error budget of this experiment the results
obtained in Sec. 2
and Sec. 3 have been used in order to calculate
the mismodeled amplitudes of the solid and ocean tidal perturbations
$\d\O^{I},\ \d\O^{II}$ and $\d\og^{II}$; they have been
subsequently  compared with the gravitomagnetic precessions over 4
years $\D\O^{I}_{LT}=124$ mas, $\D\O^{II}_{LT}=126$ mas
and $\D\og^{II}_{LT}=-228$ mas. In Tab.\ref{mismo1} we have quoted
those solid tidal lines whose mismodeled perturbative
amplitudes amount to $1\%$, at least, of the gravitomagnetic
shifts. It
turns out
that only the 18.6-year tide and $K_1$ exceed this cutoff.
Regarding the ocean tides, it turns out from Tab.\ref{mismo2} that the
perigee of LAGEOS II
is more sensitive to the mismodeling of the ocean part of the Earth
response to the tide generating potential.
In particular, the  effect of $K_1\ l=3\ p=1\ q=-1$ is relevant with
a total $\d\og=
\vass{\derp{A(\og)}{C^{+}_{lmf}}}\d C^{+}_{lmf}+
\vass{\derp{A(\og)}{k^{'}_{l}}}\d k^{'}_{l}$ of 64.5
mas
amounting to 28.3 $\%$ of $\D\og_{LT}$ over 4 years.                   

The combination $y_{LT}$ is useful since it should vanish if
calculated for the
even zonal contributions $C_{2 0}$ and $C_{4 0}$ of the geopotential [{\it Ciufolini}, 1996]. More precisely,
the right side of \rfr{resciu} should become equal to zero if the left side
were calculated for any of these two even zonal contributions, both of static and
dynamical origin; the nearer to zero is the right side, the smaller is the
systematic uncertainty in $\m$ due to the contribution considered.

In order to test preliminarily this important feature for the case of
tides, in a very
conservative way the results
obtained in the present work have been used in \rfr{resciu} assuming,
for the sake of clarity and in order to make easier
 the comparison with [{\it Ciufolini et al.},  1997], 
a 1 year time span and the nominal values of the calculated tidal
perturbative amplitudes, as if
the zonal solid and ocean tides were not at all included in the GEODYN II dynamical models so that the residuals should
account entirely for them. The results are released in Tab.\ref{ciuflt1} and Tab.\ref{ciuflt2}.      
The values of $\d\m$ quoted there for the various zonal tidal lines may
be considered as the systematic error
in $\m$ due to the chosen constituents, if considered one at a time by neglecting any possible reciprocal
correlation among the other tidal lines.     
Tab.\ref{ciuflt1} and Tab.\ref{ciuflt2} show that the percent error in the
general relativistic value of $\m$ due to
the 18.6-year tide,
the most dangerous one in recovering the LT since it superimposes to the gravitomagnetic trend over time spans of a few year,
amounts to 21.9 $\%$, while for all the other zonal tides it
reduces to 0.1 $\%$ or less. This means that, even if fully present in the combined residuals,
the $l=2\ m=0$ tides, with the exception of the 18.6-year tide, do not
affect the recovery of $\m_{LT}$ by means of $y_{LT}$ so that there is no need including
them into the final least-squares scheme. It is interesting to compare
the present results to those released in [{\it Ciufolini et al.} 1997] for the
18.6-year tide.
The value -0.219 due to the solid component for $\m$ quoted in Tab.\ref{ciuflt1} must be
compared to
-0.361 in [{\it Ciufolini et al.} 1997], with an improvement of
$39.3$ $\%$. In the cited work there is no reference to any estimate of the mismodeling of the 18.6-year tide, so that we have
used the nominal tidal perturbative amplitudes released in it:  $A(\O^{I})=-997$ mas,
$A(\O^{II})=1805$ mas and $A(\og^{II})=-1265$ mas. These figures for 
the perturbative amplitudes due to the solid Earth tide of 18.6-year are
notably different from those quoted in the present study.  In  
 [{\it Ciufolini et al.}, 1997]  the theoretical
framework in which those numbers have been calculated (F. Vespe, private
communication, 1999)
 is based on the assumption
of a spherical, static, elastic Earth with a single
nominal value of $k_2=0.317$ used  for  the 
entire tidal spectrum. The inclusion of the tiny corrections due to the
 Earth's flattening and rotation on the 
perturbative amplitudes of $\O$ and $\og$ for the 18.6-year tide 
could allow to slightly improve the related uncertainty in $\m$; it would 
amount to $20.6\ \%$. But since the present-day accuracy in laser ranging 
measurements could hardly allow to detect these small effects, their utilization
in \rfr{resciu} is debatable.

Remember that the result quoted for the 18.6-year tide
is obtained in the worst possible case, i.e. a time span of only 1 year and the
assumption that the residuals have been built up by neglecting
completely the zonal tides in the dynamical models. Recall that if the residuals accounted for
the 18.6-year tide it would not be possible to view it as an empirically fit quantity
unless a $T_{obs}$ of, at least, 18.6 years is adopted. However, if, more
realistically, we calculate \rfr{resciu}
with the mismodeled amplitudes quoted in the first row of Tab.\ref{mismo1} for the 18.6-year tide we
obtain
$\d\m=\leq -3.51\cdot10^{-3}$.
% This strongly
% highlights that many efforts, either theoretical or
%experimental, must be done in order to modeling the 
%more accurately as possible such
%a constituent so that it can be
% included in the nominal background of the orbital determination
%softwares 
%like GEODYN II at a satisfactory level of
%accuracy. The calculations performed in this work point
%toward this goal in the sense that, if we put our values for the
%perturbative
%amplitudes due to the 18.6-year tide
%in
% the models of, e.g., GEODYN II
% and subsequently build up the orbital residuals, we expect that the
%contribution of such
%semisecular constituent to $\d\m$ amounts to the value quoted here. 
This conclusion has been confirmed
also by calculating the time average of the 18.6-year tide
over different time spans by using the mismodeling level adopted here
 [{\it Pavlis and Iorio}, 2001].

Even though a cancellation is not expected as for the first two even
zonal constituents, calculating
the left hand side of \rfr{resciu} for the other tides yields, at least, an order of magnitude of their effect
on
$\m$. An interesting, unpredicted
feature stands out for the odd zonal ocean
tides. The contribution of $l=3$ zonal ocean tidal
nominal perturbations over 1 year  to $\d\m$ can be
found in Tab.\ref{ciuflt3} and Tab.\ref{ciuflt4}. 
From an inspection of them it
is clear that the sensitivity of perigee of LAGEOS II to the $l=3$ part of the ocean
tidal spectrum may affect the recovery of the Lense-Thirring parameter
$\m$ by means of the
proposed combined residuals, especially as far as $S_a$ and $S_{sa}$ are concerned. This fact agrees with the results of
Tab.\ref{mismo2} which tells us that the
mismodeled parts of $S_{a}$ and $S_{sa}$ are not negligible fractions of $\D\og^{II}_{LT}$. However, if the mismodeled amplitudes are
employed in \rfr{resciu} it can be seen that, over 1 year, a cancellation of the order of
$10^{-1}$ ($S_{a}\ l=3\ p=2$) and $10^{-2}$
($S_{a}\ l=3\ p=1;\ S_{sa}\ l=3\ p=1,2$) takes place. The contributions of the mismodeling on $M_{m}$ and $M_{f}$ are completely
negligible. So, we can conclude that also the $l=3$ part of the zonal ocean tidal spectrum may not affect the combined residuals in a
sensible manner if the $l=3$ part of  $S_a$ and $S_{sa}$ is properly accounted for.
%
%This feature, along with the assessment of the effect of the
%tesseral and sectorial tides by means of numerical simulations based on the results obtained
%here
%will be the subject of a forthcoming paper [{\it Pavlis and Iorio},.
%-----------------------------------------------
\section{Conclusions}
%In the first part of the paper an extensive, updated analysis of
%the effect of both
%solid Earth and ocean tides on the nodes $\O$ of LAGEOS and LAGEOS II
%and perigee $\og$ of LAGEOS II is carried 
%out. 
The detection of the general relativistic
Lense-Thirring drag in the field of the Earth by means of the combined
residuals
of the two LAGEOS laser-ranged satellites 
is affected,
among other factors, also by the Earth solid and ocean tidal
perturbations. 
%The calculations performed here 
%have been used, in view of a refinement of the error budget of the \lt\ experiment, in order to
%check preliminarily which tidal constituents are really important in
%perturbing the combined residuals. Tab. 5 and Tab. 9 show that, over a
%4 years time span the nodes of the two LAGEOS are sensitive to the even
%components of the $18.6$-year line, the $K_1$ and the $S_a$ at a 1$\%$
%level. Moreover, the perigee of LAGEOS II turns out to be very sensitive
%to the $l=3$ part of the ocean tidal spectrum.
% so
%that the efforts of the scientific
%community can be directed towards a better knowledge of them.
% Moreover,
%they are useful also in
%highlighting how to update the orbit determination softwares.   

Concerning the solid tides of degree $l=2$, the most effective
constituents turn out to be the semisecular 18.6-year, the $K_1$ and the
$S_2$ which induce  
on the examined orbital elements  perturbations
of the order of $10^{2}-10^{3}$ mas with periods ranging from
111.24 days for the $S_2$ on LAGEOS II  to 6798.38 days for the
18.6-year tide.

Regarding the ocean tides, we have analyzed the degree $l=2,3,4$ terms of 13 constituents by using the data of the EGM 96 model.
If, from one hand, the nominal ocean amplitudes of degree $l=2$ amount to
almost $10$$\%$ of the solid 
ones for the same degree, from the other hand the obtained results for the perigee 
of LAGEOS II show that this orbital element is very sensitive to the degree $l=3$ part of the tidal spectrum.
Indeed, for many tidal lines the 
perturbations on
it with $l=3\ p=1,2\ q=-1,1$ exhibit long periods of some years and 
amplitudes of the order of magnitude of those due to the solid Earth tides for 
the node. The $K_1\ l=3\ p=1\ q=-1$ line stands out  with its period of
-1851.9 days (5.07 years)
and amplitude of $1136$ mas.  Such term has a  $5.2$ $\%$ error. 

The calculations performed here
have been used, in view of a refinement of the error budget of the
gravitomagnetic LAGEOS experiment, in order to
check preliminarily which tidal constituents are really important in
perturbing the combined residuals so to fit and remove them from the data, if possible, or, at least,
to evaluate the systematic error induced by them. Tab. 5 and Tab. 9 show that, over a
4 years time span the nodes of the two LAGEOS are sensitive to the even
components of the $18.6$-year line, the $K_1$ and the $S_a$ at a 1$\%$
level at least. Moreover, the perigee of LAGEOS II turns out to be very sensitive
to the $l=3$ part of the ocean tidal spectrum.        

%The detection of the
%Lense-Thirring effect is affected,
%among other factors, also by the Earth solid and ocean tidal
%perturbations, as it is shown in the second part of the paper.
%The calculations performed here can
%be used in order to check which tidal constituents are really important
%in 
%perturbing the combined residuals so
%that the efforts of the scientific
%community can be directed towards a better knowledge
%of them. Moreover, they are useful also in
%highlighting how to update the orbit determination softwares.
Concerning the very long period tides, an accurate calculation from first
principles of their amplitudes, periods and initial 
phases turns out to be important in the case of an observational time span
shorter than their periods. Indeed, while for a biasing true linear
trend one can only assess as more accurately as possible the error induced by it, in the
case of the arc of sinusoid of a harmonic perturbation whose
period is longer than the adopted time span it should be possible to
fit and remove it from the signal, without affecting the trend of
interest, provided that its period and
initial phases are exactly known. In any case it could be possible to assess reliably
its role in the error budget. 

However, in the
Lense-Thirring experiment we have shown in a preliminary way that this should not be necessary, at least for the
semisecular even zonal 18.6-year tide independently of the time series
length.
Indeed, if calculated at the level
of accuracy shown in this paper, it affects the combined
residuals at a level $\leq 10^{-3}$. The other even zonal tides do not
create problems. Also the $l=3\ m=0$ tides, and this is an
unpredicted feature cancel out at a
level $\leq 10^{-1}$-$10^{-2}$ if they are properly accounted for in 
building up the residuals. The results
presented in this paper not only confirm the
usefulness of the formula by Ciufolini
in canceling out the $l=2,4\ m=0$ tides, but also extend
its validity to the $l=3\ m=0$ part of the tidal response spectrum.

The impact of the tesseral ($m=1$) and sectorial ($m=2$) tidal constituents on the combined residuals, which are fully sensitive to them,
over different time spans and time steps will be the subject of a forthcoming numerical analysis based on
the results presented here. 
\vspace*{1cm}

\noindent{\bf Acknowledgments\\}
\noindent I am thoroughly grateful to Prof. I. Ciufolini who caught my attention to gravitomagnetism and, in particular, suggested me to
address the problem of the influence of Earth tides on the Lense-Thirring effect. I wish to thank also Prof. L. Guerriero who
supported me here
at Bari. I thank also E. C.  Pavlis, S. V. Bettadpur, F. Vespe, R. Eanes, V. Dehant, L. Petrov, H. Schuh for the useful material
sent
to me and fruitful discussions.

\onecolumn
\begin{table}[ht!] \centering\footnotesize{\btab{||c||c|c|c|c|c||} \hline \multicolumn{6}{||c||}{Solid
Earth tidal perturbations on the
node $\O$
 of LAGEOS}\\
\hline
Tide & \multicolumn{5}{c||} { l=2, p=1, q=0} \\ \cline{2-6}
      & Period (days) & Amplitude (mas) & $k_2$ Love number &
$H_{l}^{m}(f)$ (meters)&
$\tan{\d_{lmf}}$\\
\hline
055.565 & 6798.38 & -1079.38 & 0.315 & 0.02792 & -0.01715 \\ \hline
055.575 & 3399.19 & 5.23 & 0.313 & 0.000272 & -0.015584 \\ \hline
056.554 $S_a$ & 365.27 & 9.96 & 0.307 & -0.00492 & -0.01135 \\ \hline
057.555 $S_{sa}$ & 182.62 & 31.21 & 0.305 & -0.03099 & -0.01029 \\ \hline
065.455 $M_m$ & 27.55 & 5.28 & 0.302 & -0.03518 & -0.00782 \\ \hline
075.555 $M_{f}$ & 13.66 & 4.94 & 0.301 & -0.06659 & -0.007059 \\
\hline\hline
165.545 & 1232.94 & -41.15 & 0.259 & -0.007295 & -0.00554\\ \hline
165.555 $K_1$& 1043.67 & 1744.38 & 0.257 & 0.3687012 & -0.0055933 \\
\hline
165.565 & 904.77 & 203.02 & 0.254 & 0.050028 & -0.005653\\ \hline
163.555 $P_1$ & -221.35 & 136.44 & 0.286 & -0.12198 & -0.005017\\ \hline
145.555 $O_1$& -13.84& 19 & 0.297 & -0.26214 & -0.00484 \\ \hline
135.655 $Q_1$& -9.21& 2.42& 0.297 & -0.05019 & -0.00483 \\ \hline\hline
274.556 & -1217.55 & 1.68 & 0.301 & 0.000625 & -0.00431 \\ \hline
274.554 & -1216.73 & -6.63 & 0.301 & -0.00246 & -0.00431\\ \hline
275.555 $K_2$ & 521.835 & -92.37 & 0.301 & 0.0799155 & -0.00431 \\ \hline
273.555 $S_2$& -280.93 & 182.96 & 0.301 & 0.2940 & -0.00431 \\ \hline
272.556 $T_2$& -158.80 & 6.04 & 0.301 & 0.0171884 & -0.00431 \\ \hline
255.555 $M_2$& -14.02 & 19.63 & 0.301 & 0.6319 & -0.00431\\ \hline
245.655 $N_2$& -9.29 & 2.49 & 0.301 & 0.12099 & -0.00431\\ \hline
\etab}
\caption{\footnotesize{Perturbative amplitudes on the node $\O$ of LAGEOS
due   
 to solid Earth
tides. In the first column the Doodson number of each constituent is
quoted
followed by the Darwin's name, when it is present. The tidal lines are
listed
in order of decreasing periods. The coefficients
$H_l^m(f)$ are those recently calculated by Roosbeek
and multiplied by suitable normalization factors (IERS standards) 
in order to make possible a comparison with those of
Cartwright and Edden. $\tan{\d_{lmf}}$ expresses the phase
lag of the solid Earth
response with respect to the tidal potential due to the anelasticity in
the
mantle.}}
\label{uno}
\end{table}

\begin{table}[ht!] \centering
\footnotesize{\btab{||c||c|c|c|c|c||} \hline
\multicolumn{6}{||c||}{Solid Earth  tidal perturbations on the node $\O$
 of LAGEOS II}\\
\hline
Tide & \multicolumn{5}{c||} { l=2, p=1, q=0} \\ \cline{2-6}
      & Period  (days)& Amplitude  (mas)& $k_2$ Love number
& $H_{l}^{m}(f)$ (meters)&
$\tan{\d_{lmf}}$\\
\hline   
055.565  & 6798.38  & 1982.16  & 0.315 & 0.02792  & -0.01715 \\ \hline   
055.575  & 3399.19 & -9.61 & 0.313 & -0.000272 & -0.015584 \\ \hline
056.554 $S_a$  & 365.27  &  -18.28  & 0.307 & -0.00492  & -0.01135 \\
\hline
057.555 $S_{sa}$ & 182.62   & -57.31  & 0.305 & -0.03099  & -0.01029 \\
\hline
065.455 $M_m$  & 27.55  & -9.71 &  0.302 & -0.03518 & -0.00782 \\ \hline
075.555 $M_f$  &  13.66 & -9,08   & 0.301 & -0.06659   & -0.007059 \\
\hline\hline
165.565  & -621.22   & -58.31  & 0.254 & 0.050028  & -0.005653 \\ \hline
165.555  $K_1$& -569.21  &-398  & 0.257 & 0.3687012  & -0.0055933 \\
\hline
165.545  & -525.23  & 7.33  & 0.259 & -0.007295  & -0.005541 \\ \hline
163.555  $P_1$& -138.26  & 35.65  & 0.286 & -0.1219  & -0.005017 \\ \hline
145.555  $O_1$&-13.33& 7.66 & 0.297 & -0.26214  & -0.00484 \\ \hline
135.655  $Q_1$& -8.98& 0.98 & 0.297 & -0.05019  & -0.00483  \\
\hline\hline
275.555  $K_2$& -284.6 &-92.51  & 0.301 & 0.079915  & -0.004318 \\ \hline
274.556  &-159.96  &-0.40  & 0.301 & 0.000625  & -0.00431 \\ \hline
274.554  & -159.95 & 1.6  & 0.301 & -0.00246  & -0.0043 \\ \hline
273.555  $S_2$&-111.24  & -133.04  & 0.301 & 0.29402  & -0.00431 \\ \hline
272.556 $T_2$& -85.27 & -5.96 & 0.301 & 0.017188 & -0.00431\\ \hline
255.555  $M_2$&-13.03 & -33.05 & 0.301 & 0.6319  & -0.004318 \\ \hline
245.655  $N_2$&-8.84 & -4.35 & 0.301 & 0.12099  & -0.00431  \\ \hline
\etab}
\caption{\footnotesize{Perturbative
 amplitudes on the node $\O$ of LAGEOS II due to solid Earth
tides. In the first column the Doodson number of each constituent is
quoted
followed by the Darwin's name, when it is present. The tidal lines are   
listed
in order of decreasing periods. The coefficients $H_l^m(f)$ are those
recently calculated by Roosbeek
and multiplied by suitable normalization factors (IERS standards)   
in order to make possible a comparison with those of
Cartwright and Edden. $\tan{\d_{lmf}}$ expresses the phase lag of the   
solid Earth
response with respect to the tidal potential due to the anelasticity in
the
mantle.}}
\label{due}
\end{table}

\begin{table}[ht!] \centering
\footnotesize{\btab{||c||c|c|c|c|c||} \hline
\multicolumn{6}{||c||}{Solid Earth  tidal perturbations on the perigee
$\og$
 of LAGEOS II}\\
\hline
Tide & \multicolumn{5}{c||} { l=2, p=1, q=0} \\ \cline{2-6}
      & Period (days) & Amplitude (mas) & $k_2$ Love number
& $H_{l}^{m}(f)$ (meters)&
$\tan{\d_{lmf}}$\\
\hline
055.565  & 6798.38  & -1375.58  & 0.315 & 0.02792  & -0.01715 \\ \hline
055.575  & 3399.19 & 6.66 & 0.313 & -0.000272 & -0.015584 \\ \hline
056.554 $S_a$  & 365.27  &  12.69  & 0.307 & -0.00492  & -0.01135 \\
\hline   
057.555 $S_{sa}$ & 182.62   & 39.77  & 0.305 & -0.03099  & -0.01029 \\   
\hline
065.455 $M_m$  & 27.55  & 6.74  &  0.302 & -0.03518 & -0.00782 \\ \hline
075.555 $M_f$  &  13.66 & 6.30  & 0.301 & -0.06659   & -0.007059 \\
\hline\hline
165.565  & -621.22   & 290.43  & 0.254 & 0.050028  & -0.005653 \\ \hline
165.555  $K_1$& -569.21  & 1982.14 & 0.257 & 0.3687012  & -0.0055933 \\ 
\hline
165.545  & -525.23  & -36.52  & 0.259 & -0.007295  & -0.005541 \\ \hline
163.555  $P_1$& -138.26  & -177.56  & 0.286 & -0.1219  & -0.005017 \\
\hline   
145.555  $O_1$&-13.33 & -38.16 & 0.297 & -0.26214  & -0.00484 \\ \hline
135.655  $Q_1$& -8.98& -4.92 & 0.297 & -0.05019  & -0.00483  \\
\hline\hline
275.555  $K_2$& -284.6 & -88.19 & 0.301 & 0.079915  & -0.004318 \\ \hline
274.556  &-159.96  &-0.38  & 0.301 & 0.000625  & -0.00431 \\ \hline
274.554  & -159.95 & 1.52  & 0.301 & -0.00246  & -0.0043 \\ \hline
273.555  $S_2$&-111.24  &-126.83   & 0.301 & 0.29402  & -0.00431 \\ \hline
272.556 $T_2$& -85.27 & -5.68 & 0.301 & 0.017188 & -0.00431\\ \hline
255.555  $M_2$&-13.03& -31.9 & 0.301 & 0.6319  & -0.004318 \\ \hline
245.655  $N_2$&-8.84&-4.15 & 0.301 & 0.12099  & -0.00431  \\ \hline
\etab}
\caption{\footnotesize{Perturbative
 amplitudes on the perigee $\og$ of LAGEOS II due to solid Earth
tides. In the first column the Doodson number of each constituent is
quoted
followed by the Darwin's name, when it is present. The tidal lines are
listed
in order of decreasing periods. The coefficients $H_l^m(f)$ are those
recently calculated by Roosbeek
and multiplied by suitable normalization factors (IERS standards)
in order to make possible a comparison with those of
Cartwright and Edden. $\tan{\d_{lmf}}$ expresses the phase lag of the
solid Earth 
response with respect to the tidal potential due to the anelasticity in 
the
mantle.}}
\label{trec}
\end{table}

\begin{table}[h!] \centering
\footnotesize{\btab{||c||c|c|c|c||} \hline
\multicolumn{5}{||c||}{Corrections due to Earth's flattening and rotation to 
solid tidal perturbations}\\
\hline
Tide & \multicolumn{4}{c||} { l=2,  p=2, q=0} \\ \cline{2-5}
      & $\D\O_{LAGEOS}$
 (mas)& $\D\O_{LAGEOS II}$ (mas) & $\D\og_{LAGEOS II}$ (mas)& 
$k_{2m}^{+}$\\
\hline 
055.565  & 0.43 & 1.28 & 0.19 & -0.00094 \\ \hline
165.555 $K_1$ & -0.97 & -9.49 & 0.85 & -0.00074 \\ \hline
\etab}
\caption{\footnotesize{Corrections to the Earth solid tidal perturbations on 
the nodes of LAGEOS and the perigee of LAGEOS II due to the Earth's flattening 
and Earth's rotation.
The values adopted for $k_{2m}^{+}$ are those quoted by {\it Dehant}
[1999].}}
\label{k+}      
\end{table}

\begin{table}[ht!] \centering \tiny{\btab{||c||c||c|c||c|c||c|c||} \hline \multicolumn{8}{||c||}{ Mismodeled solid tidal
perturbations on nodes
of LAGEOS and LAGEOS II and perigee of LAGEOS II}\\
\hline
Tide & \multicolumn{7}{c||} { $\D\O^{I}_{LT}=124$ mas \ $\D\O^{II}_{LT}=126$ mas \ $\D\og^{II}_{LT}=-228$ mas \ $P_{obs}$=4 years} \\
\cline{2-8}
      & $\d k_2/k_2$ ($\%$) & $\d\O^{I}$ (mas) & $\d\O^{I}/\D\O^{I}_{LT}$ ($\%$) & $\d\O^{II}$ (mas) &
$\d\O^{II}/\D\O^{II}_{LT}$ ($\%$) & $\d\og^{II}$ (mas) & $\d\og^{II}/\D\og^{II}_{LT}$ ($\%$)\\
\hline
055.565 & 1.5 & -16.5 & 13.3 & 30.3 & 24 & -21 & 9.2 \\ \hline
165.555 $K_1$& 0.5 & 9 & 7.2 & -2 & 1.6 & 10.2 & 4.4 \\
\hline
\etab}
\caption{\footnotesize{Mismodeled solid tidal perturbations on nodes $\O$ of LAGEOS  and LAGEOS II and the perigee $\og$ of
LAGEOS II compared to their gravitomagnetic precessions over 4
years.}}
\label{mismo1}
\end{table}

\begin{table}[h!]
 \centering \tiny{\btab{||c||c|c|c||c|c|c||c|c|c||c|c|c||} \hline
\multicolumn{13}{||c||}{Ocean tidal perturbations on the node $\O$ of LAGEOS}\\
\hline
Tide & \multicolumn{3}{c||} { l=2, p=1, q=0} & \multicolumn{3}{c||}
 { l=3, p=1, q=-1} & \multicolumn{3}{c||} { l=3, p=2, q=1} &
\multicolumn{3}{c||} { l=4, p=2, q=0}\\ \cline{2-13}
      & P  & A  & E & P  & A & E  
      & P  & A & E & P  & A & E\\
\hline 
065.455 Mm & 27.55   & -0.54 & 14.4  & 28 & $-10^{-4}$ & 66.6  & 27.11 &
 $10^{-4}$ &
 66.6  & - & - & - \\
\hline
056.554 Sa & 365.27 & -20.55 & 6.7 & 464.67 & $-10^{-2}$ & 10 & 300.91 &
 $10^{-2}$ & 10 
& - & - & -\\ \hline
075.555 Mf & 13.66 & -0.62 & 7.8 & 13.77 & $-10^{-4}$ & 112 & 13.55 & $10^{-4}$
  & 112 & - 
& - & -\\ \hline
057.555 Ssa & 182.62 & -5.98 & 9.4 & 204.5 & $-10^{-3}$ & 27.2 & 164.9 &
 $10^{-3}$  & 27.2
& - & - & -\\ \hline \hline
165.555 K1 & 1043.67 & 156.55 & 3.8 & 2684.2 & -0.36 & 5.2 & 647.76 & 
$10^{-2}$  & 5.2 & 1043.67 & 4.63 & 3.9\\ \hline
163.555 P1 & -221.35 & -11.49 & 8.1 & -195.95 & $10^{-3}$ & 18.5 & -254.3 &
 $-10^{-3}$ & 
18.5 & -221.35 & -0.32 & 8\\ \hline
145.555 O1 & -13.84 & -2 & 2.9 & -13.72 & $10^{-3}$ & 3.2 & -13.95 & 
$10^{-4}$ 
& 3.2 & -13.84 & $10^{-2}$ & 5.7 \\ \hline
135.655 Q1 & -9.21 & -0.28 & 13.5 & -9.16 & $10^{-4}$ & 25
 & -9.26 & $10^{-5}$ & 25 & -9.21 &
$-10^{-3}$ & 20\\ \hline\hline 
275.555 K2 & 521.83 & -6.24 & 11.1 & 751.5 & $-10^{-2}$ & 5.5 & 399.7 &
 $10^{-2}$  & 5.5 &
521.83 & -9.58 & 15.4\\ \hline
273.555 S2 & -280.93 & 9.45 & 3.9 & -241.24 & $10^{-3}$  & 7.1 & -336.25 &
 $-10^{-2}$  & 7.1 &
-280.93 & 15.08 & 5.2\\ \hline
272.556 T2 & -158.8 & 0.28 & 75 & -145.3 & $10^{-4}$  & 50 & -175 & 
$-10^{-3}$  & 50 &
-158.8 & 0.44 & 100 \\ \hline
255.555 M2 & -14.02 & 2.03 & 0.9 & -13.9 & $-10^{-4}$ & 7.4 & -14.14 &
 $10^{-3}$  & 7.4 &
-14.02 & 2.08 & 2.8\\ \hline
245.655 N2 & -9.29 & 0.3 & 4.6 & -9.2 &  $10^{-5}$ & 12.5 & -9.3 & 
$10^{-4}$  & 12.5 
& -9.29 & 0.3 & 8.3\\ \hline
\etab}
\caption{\footnotesize{Perturbative
 amplitudes on the node $\O$ of LAGEOS due to ocean 
tides. P indicates the periods in days, A the amplitudes in mas and E the 
percent error in the $C_{lmf}^{+}$. The values employed for them and the 
related errors are those quoted in EGM96 model.}}
\label{otto}      
\end{table}
        
\begin{table}[h!]
 \centering \tiny{\btab{||c||c|c|c||c|c|c||c|c|c||c|c|c||} \hline
\multicolumn{13}{||c||}{Ocean tidal perturbations on the node $\O$ of LAGEOS 
II}\\
\hline
Tide & \multicolumn{3}{c||} { l=2, p=1, q=0} & \multicolumn{3}{c||}
 { l=3, p=1, q=-1} & \multicolumn{3}{c||} { l=3, p=2, q=1} &
\multicolumn{3}{c||} { l=4, p=2, q=0}\\ \cline{2-13}
      & P  & A  & E & P  & A & E  
      & P  & A & E & P  & A & E\\
\hline 
065.455 Mm  & 27.55  &  1  &  14.4  &  26.65  &  $10^{-3}$  &  66.6  &
 28.5  &  -$10^{-3}$ & 66.6  &  - & - & - \\
\hline 
056.554 Sa & 365.27 & 37.71 & 6.7 & 252.8 & 0.13 & 10 & 657.5 & -0.35 & 10 & - 
& - & -\\ \hline
075.555 Mf & 13.6 & 1.13 & 7.8 & 13.43 & $10^{-4}$ & 112 & 13.89 & $-10^{-4}$
 & 112 & - & 
- & - \\ \hline
057.555 Ssa & 182.62 & 10.98 & 9.4 & 149.41 & $10^{-2}$ & 27.2 & 234.8 &
 $-10^{-2}$ & 27.2 & - & - & -\\ \hline \hline
165.555 K1 & -569.21 & -35.69 & 3.8 & -1851.9 & -1.02 & 5.2 & -336.3 & $-10
^{-3}$ &
5.2 & -569.21 & 41.58 & 3.9 \\ \hline
163.555 P1 & -138.26 & -3 & 8.1 & -166.23 & $-10^{-2}$ & 18.5 & -118.35 & 
$-10^{-4}$ & 18.5 & -138.26 & 3.29 & 8 \\ \hline
145.555 O1 & -13.3 & -0.8 &  2.9 & -13.5 & -$10^{-2}$ & 3.2 &  -13.12 & 
$-10^{-4}$ & 3.2 & -13.3 & 0.7 & 5.7 \\ \hline
135.655 Q1 & -8.98 & -0.11 & 13.5 & -9.08 & -$10^{-3}$ & 25 & -8.89  & 
$10^{-5}$ 
& 25 & -8.98 & 0.11 & 20 \\ \hline\hline
275.555 K2 & -284.6 & -6.24 & 11.1 & -435.38 & -0.13 & 5.5 & -211.4 & 
$10^{-2}$ &
5.5 & -284.6 & -5.95 & 15.4 \\ \hline
273.555 S2 & -111.2 & -6.87 & 3.9 & -128.6 & $-10^{-2}$ & 7.1 & -97.9 &
 $10^{-2}$ & 
7.1 & -111.2 & -6.79 & 5.2 \\ \hline
272.556 T2 & -85.27 & -0.277 & 75 & -95.14 & $-10^{-3}$ &  50 &
 -77.25 & $10^{-3}$ & 50 & -85.27 & -0.274 & 100 \\ \hline
255.555 M2 & -13.03 & -3.46 & 0.9 & -13.2 & -$10^{-3}$ & 7.4 & -12.83
 & $10^{-3}$ 
& 7.4 & -13.03 & -2.2 & 2.8 \\ \hline
245.655 N2 & -8.8 & -0.46 & 4.6 & -8.9 & $-10^{-3}$ & 12.5 & -8.7 & 
$10^{-4}$ & 12.5 & -8.8 & -0.34 & 8.3 \\ \hline
\etab}
\caption{\footnotesize{Perturbative
 amplitudes on the node $\O$ of LAGEOS II due to ocean 
tides. P indicates the periods in days, A the amplitudes in mas
 and E the 
percent error in the $C_{lmf}^{+}$. The values employed for them and the 
related errors are those quoted in EGM96 model.}}
\label{nove}      
\end{table}        

\begin{table}[h!]
 \centering \tiny{\btab{||c||c|c|c||c|c|c||c|c|c||c|c|c||} \hline
\multicolumn{13}{||c||}{Ocean tidal perturbations on the perigee $\omega$
 of LAGEOS II}\\
\hline
Tide & \multicolumn{3}{c||} { l=2, p=1, q=0} & \multicolumn{3}{c||}
 { l=3, p=1, q=-1} & \multicolumn{3}{c||} { l=3, p=2, q=1} &
\multicolumn{3}{c||} { l=4, p=2, q=0}\\ \cline{2-13}
      & P  & A  & E & P  & A & E  
      & P  & A & E & P  & A & E\\
\hline 
065.455 Mm & 27.55 & -0.69 & 14.4 & 26.65 & -1.53 & 66.6 & 28.5 & 1.64  & 66.6
 & - & - &- 
\\ \hline
056.554 Sa & 365.27 & -26.17 & 6.7 & 252.8 & -114.35 & 10 & 657.55 & 297.34
 & 10 & 
- & - & - \\ \hline
075.555 Mf & 13.66 & -0.78 & 7.8 & 13.43 & -0.58 & 112 & 13.89 & 0.60 & 112 & - 
& - & - \\ \hline
057.555 Ssa & 182.6 & -7.62 & 9.4 & 149.41 & -22.95  & 27.2  & 234.8 & 36.07 &
27.2 & - & - & - \\ \hline \hline
165.555 K1 & -569.21 & 177.76 & 3.8 & -1851.9 & -1136 & 5.2 & -336.28 & 346.6
 & 5.2 & -569.21 & -3.95 & 3.9 \\ \hline
163.555 P1 & -138.26 & 14.95 & 8.1 & -166.2 & -28.97 & 18.5 & -118.35 & 34.67
 & 18.5 & -138.2 & -0.31 & 8 \\ \hline
145.555 O1 & -13.3 & 4 & 2.9 & -13.55 & -13.7 & 3.2 &
 -13.12 & 22.3
& 3.2 & -13.3 & -$10^{-2}$ & 5.7 \\ \hline
135.655 Q1 & -8.98 & 0.58 & 13.5 & -9.08 & -1.17 & 25 & -8.89 & 1.92
 & 25 
& -8.98 & -$10^{-2}$ &  20 \\ \hline\hline
275.555 K2 & -284.6 & -5.95 & 11.1 & -435.3 & 214.23 & 5.5 & -211.4 & 87.3
 & 5.5 &
-284.6 & -2.49 & 15.3 \\ \hline
273.555 S2 & -111.2 & -6.55 & 3.9 & -128.6 &  98.47 & 7.1 & -97.9 & 62.9
 & 7.1 & 
-111.2 & -2.85 & 5.2 \\ \hline
272.556 T2 & -85.2 & -0.26 & 75 & -95.1 & 5.2 & 50 & -77.2 & 3.54 & 50 & -85.2
 &
-0.11 & 100 \\ \hline
255.555 M2 & -13.03 & -3.3 & 0.9 & -13.2 & 9.7 & 7.4 & -12.83 & 7.9
 & 7.4 & 
-13.03 & -0.92 & 2.8 \\ \hline
245.655 N2 & -8.48 & -0.44 & 4.6 & -8.94 & 1.95 & 12.5 & -8.75 & 1.6
 & 12.5 
& -8.84 & -0.14 & 8.3 \\ \hline
\etab}
\caption{\footnotesize{Perturbative amplitudes
 on the perigee $\omega$ of LAGEOS II due to
 ocean 
tides. P indicates the periods in days, A the amplitudes in mas and E the 
percent error in the $C_{lmf}^{+}$. The values employed for them and the 
related errors are those quoted in EGM96 model.}}
\label{dieci}      
\end{table}

\begin{table}[ht!] \centering \tiny{\btab{||c||c||c|c||c|c||c|c||} \hline \multicolumn{8}{||c||}{ Mismodeled ocean tidal
perturbations on nodes
of LAGEOS and LAGEOS II and perigee of LAGEOS II}\\
\hline
Tide & \multicolumn{7}{c||} { $\D\O^{I}_{LT}=124$ mas  $\D\O^{II}_{LT}=126$ mas  $\D\og^{II}_{LT}=-228$ mas  $P_{obs}=4$ years} \\
\cline{2-8}
      & $\d C^{+}/C^{+}$ ($\%$)  & $\d\O^{I}$ (mas) & $\d\O^{I}/\D\O^{I}_{LT}$ ($\%$)  & $\d\O^{II}$ (mas) &
$\d\O^{II}/\D\O^{II}_{LT}$ ($\%$) & $\d\og^{II}$  (mas)& $\d\og^{II}/\D\og^{II}_{LT}$ ($\%$) \\
\hline
$S_{a}$ $l$=2 $p$=1 $q$=0& 6.7 & 1.37 & 1.1 & 2.5 & 1.9 & - & - \\ \hline
$S_{a}$ $l$=3 $p$=1 $q$=-1 & 10 & - & - & - & - & 11.4 & 5 \\ \hline
$S_{a}$ $l$=3 $p$=2 $q$=1 & 10 & - & - & - & - & 29.7 & 13 \\ \hline\hline
$S_{sa}$ $l$=3 $p$=1 $q$=-1 & 27.2 & - & - & - & - & 6.2 & 2.7 \\ \hline   
$S_{sa}$ $l$=3 $p$=2 $q$=1 & 27.2 & - & - & - & - & 9.8 & 4.3 \\ \hline\hline
$K_1$ $l$=2 $p$=1 $q$=0& 3.8 & 5.9 & 4.7 & 1.3 & 1 & 6.75& 2.9\\ \hline
$K_1$ $l$=3 $p$=1 $q$=-1& 5.2 & - & - & - & - & 64.5 & 28.3\\ \hline
$K_1$ $l$=3 $p$=2 $q$=1& 5.2 & - & - & - & - & 18 & 7.9\\ \hline   
$K_1$ $l$=4 $p$=2 $q$=0& 3.9 & - & - & 1.6 & 1.2 & - & -\\ \hline\hline
$P_1$ $l$=3 $p$=1 $q$=-1& 18.5 & - & - & - & - & 5.3 & 2.3\\ \hline
$P_1$ $l$=3 $p$=2 $q$=1& 18.5 & - & - & - & - & 6.4 & 2.8\\ \hline\hline
$K_2$ $l$=3 $p$=1 $q$=-1& 5.5 & - & - & - & - & 11.7 & 5\\ \hline
$K_2$ $l$=3 $p$=2 $q$=1& 5.5 & - & - & - & - & 4.8 & 2\\ \hline\hline
$S_2$ $l$=3 $p$=1 $q$=-1& 7.1 & - & - & - & - & 6.9 & 3\\ \hline
$S_2$ $l$=3 $p$=2 $q$=1& 7.1 & - & - & - & - & 4.4 & 1.9\\ \hline\hline
$T_2$ $l$=3 $p$=1 $q$=-1& 50 & - & - & - & - & 2.6 & 1.1\\ \hline
\etab}
\caption{\footnotesize{Mismodeled ocean tidal perturbations on nodes $\O$ of LAGEOS  and LAGEOS II and the perigee $\og$ of
LAGEOS II compared to their gravitomagnetic precessions over 4
years. The effect of the ocean loading has been neglected.  When the $1\%$ cutoff has
not been reached a - has been inserted. The values quoted for $K_1\ l=3\ p=1$ includes also the mismodeling in the ocean loading coefficient
$k^{'}_{3}$ assumed equal to $2\%$.}}
\label{mismo2}
\end{table}

\begin{table}[ht!] \centering
\btab{||c||c|c|c|c||} \hline
\multicolumn{5}{||c||}{Solid Earth even zonal tidal contributions to $\d\m$}\\
\hline
Tide & \multicolumn{4}{c||} { l=2, m=0, p=1, q=0} \\ \cline{2-5}
      & $A(\O_{I})$
 (mas)& $A(\O_{II})$ (mas) & $A(\og_{II})$ (mas)&
$\d\m$\\
\hline
055.565 & -1079.38 & 1982.16 & -1375.58 & -0.219 \\ \hline
055.575 & 5.23 & -9.61 & 6.66 & $1.06\cdot 10^{-3}$\\ \hline
056.554 $S_a$ & 9.95 & -18.28 & 12.69 & $2.02\cdot 10^{-3}$ \\ \hline
057.555 $S_{sa}$ & 31.21 & -57.31 & 39.77 & $6.33\cdot 10^{-3}$ \\ \hline
065.455 $M_m$ & 5.28 & -9.71 & 6.74 & $1.07\cdot 10^{-3}$ \\ \hline
075.555 $M_{f}$ & 4.94 & -9.08 & 6.3 & $1\cdot 10^{-3}$ \\ \hline
\etab
\caption{\footnotesize{Contribution of
the even zonal solid tidal constituents to
$\d\m$ by means of the formula $\d\dot\O^{I}+\d\dot\O^{II}\times 0.295-
\d\dot\og^{II}\times 0.35=60.05\times \m$.}}
\label{ciuflt1}
\end{table}

\begin{table}[ht!] \centering
\btab{||c||c|c|c|c||} \hline
\multicolumn{5}{||c||}{Ocean even zonal tidal contributions to $\d\m$}\\
\hline
Tide & \multicolumn{4}{c||} { l=2, m=0, p=1, q=0} \\ \cline{2-5}
      & $A(\O_{I})$
 (mas)& $A(\O_{II})$ (mas) & $A(\og_{II})$ (mas)&
$\d\m$\\
\hline
056.554 Sa & -20.55 & 37.71 & -26.17 & $-3.68\cdot 10^{-3}$ \\ \hline
057.555 Ssa & -5.98 & 10.98 & -7.62 & $-1.28\cdot 10^{-3}$ \\ \hline
065.455 Mm & -0.54 & 1 & -0.69 & $-8.78\cdot 10^{-5}$ \\ \hline
075.555 Mf & -0.62 & 1.13 & -0.78 & $-1.73\cdot 10^{-4}$ \\ \hline
\etab   
\caption{\footnotesize{Contribution of the even zonal ocean tidal
 constituents to
$\d\m$ by means of the formula $\d\dot\O^{I}+\d\dot\O^{II}\times 0.295-
\d\dot\og^{II}\times 0.35=60.05\times \m$.}}
\label{ciuflt2}
\end{table}

\begin{table}[ht!] \centering
\btab{||c||c|c|c|c||} \hline
\multicolumn{5}{||c||}{Ocean odd zonal tidal contributions to $\d\m$}\\
\hline
Tide & \multicolumn{4}{c||} { l=3, m=0, p=1, q=-1} \\ \cline{2-5}
      & $A(\O_{I})$
 (mas)& $A(\O_{II})$ (mas) & $A(\og_{II})$ (mas)&
$\d\m$\\
\hline
056.554 Sa & -0.063 & 0.13 & -114.35 & $0.66$ \\ \hline
057.555 Ssa & $-9\cdot 10^{-3}$ & 0.028 & -22.95 & $0.133$ \\ \hline   
065.455 Mm & $-4\cdot 10^{-4}$ & $1\cdot 10^{-3}$ & -1.53 & $-8.93\cdot 10^{-3}$ \\ \hline
075.555 Mf & $-1\cdot 10^{-4}$ & $7\cdot 10^{-4}$ & -0.58 &
 $3.41\cdot 10^{-3}$ \\ \hline
\etab
\caption{\footnotesize{Contribution of the odd
 zonal ocean tidal constituents to
$\d\m$ by means of the formula $\d\dot\O_{I}+\d\dot\O^{II}\times 0.295-
\d\dot\og^{II}\times 0.35=60.05\times \m$ for  $p=1,\ q=-1$.}}
\label{ciuflt3}
\end{table}

\begin{table}[ht!] \centering
\btab{||c||c|c|c|c||} \hline
\multicolumn{5}{||c||}{Ocean odd zonal tidal contributions to $\d\m$}\\
\hline
Tide & \multicolumn{4}{c||} { l=3, m=0, p=2, q=1} \\ \cline{2-5}
      & $A(\O_{I})$
 (mas)& $A(\O_{II})$ (mas) & $A(\og_{II})$ (mas)&
$\d\m$\\
\hline
056.554 Sa & 0.047 & -0.36 & 297.34 & $-1.72$ \\ \hline
057.555 Ssa & $7\cdot 10^{-3}$ & -0.044 & 36.07 & $-0.209$ \\ \hline
065.455 Mm & $4\cdot 10^{-4}$ & $-2\cdot 10^{-3}$ & 1.642 & $-9.55\cdot 10^{-3}$ \\ \hline
075.555 Mf & $1.79\cdot 10^{-4}$ & $-7\cdot 10^{-4}$ & -0.60 &
 $-3.52\cdot 10^{-3}$ \\ \hline
\etab   
\caption{\footnotesize{Contribution of the odd
 zonal ocean tidal constituents to
$\d\m$ by means of the formula $\d\dot\O^{I}+\d\dot\O^{II}\times 0.295-
\d\dot\og^{II}\times 0.35=60.05\times \m$ for  $p=2,\ q=1$.}}
\label{ciuflt4}
\end{table}
\end{document}